%% file: main.tex
\documentclass[10pt, conference, compsocconf]{IEEEtran}
\IEEEoverridecommandlockouts
\def\BibTeX{{\rm B\kern-.05em{\sc i\kern-.025em b}\kern-.08em
    T\kern-.1667em\lower.7ex\hbox{E}\kern-.125emX}}
\usepackage{cite}
\usepackage{amsmath,amssymb,amsfonts}
\usepackage{algorithmic}
\usepackage{graphicx}
\usepackage{textcomp}
\usepackage{xcolor}
\usepackage{xspace}
\usepackage{enumitem,kantlipsum}
\usepackage{multirow}
\usepackage{pifont}

\newcommand{\hanqing}[1]{{\color{black} #1}}
\usepackage{subfigure}
\usepackage{url}

\begin{document}
\newcommand{\ours}{NEC\xspace}
\title{\ours: Speaker Selective Cancellation via Neural Enhanced Ultrasound Shadowing}

\author{
  \IEEEauthorblockN{%
    Hanqing Guo\IEEEauthorrefmark{1},
    Chenning Li\IEEEauthorrefmark{1},
    Lingkun Li,
    Zhichao Cao,
    Qiben Yan,
    Li Xiao%
  }%
  \IEEEauthorblockA{Department of Computer Science and Engineering,  Michigan State University}%
  \IEEEauthorblockA{\{guohanqi, lichenni, lilingk1, caozc, qyan, lxiao\}@msu.edu}\
  \IEEEauthorblockA{\IEEEauthorrefmark{1}These authors contributed equally}%

}

\definecolor{blue}{HTML}{0024ff}
\newcommand{\rev}[1]{{#1}}

\definecolor{green}{HTML}{3049D4}
\newcommand{\cl}[1]{{\color{green} \sf (cl: #1)}}


\maketitle
\input{0_abstract}
\input{1_introduction}

\input{2_related_work}

\input{3_0_background}

\input{3_system_design}

\input{4_evaluation}

\input{5_discussion}

\input{6_conclusion}

\input{7_aknowledgment}

\newpage

\input{refs}
\end{document}

%% file: 0_abstract.tex
\begin{abstract}

In this paper, we propose \ours (\underline{N}eural \underline{E}nhanced \underline{C}ancellation), a defense mechanism, which prevents unauthorized microphones from capturing a target speaker's voice. 
Compared with the existing scrambling-based audio cancellation approaches, \ours can selectively remove a target speaker's voice from a mixed speech without causing interference to others. 
Specifically, for a target speaker, we design a Deep Neural Network (DNN) model to extract high-level speaker-specific but utterance-independent vocal features from his/her reference audios. When the microphone is recording, the DNN generates a shadow sound to cancel the target voice in real-time. 
Moreover, we modulate the audible shadow sound onto an ultrasound frequency, making it inaudible for humans. By leveraging the non-linearity of the microphone circuit, the microphone can accurately decode the shadow sound for target voice cancellation. 
We implement and evaluate \ours comprehensively with 8 smartphone microphones in different settings.
The results show that \ours effectively mutes the target speaker at a microphone
without interfering with other users' normal conversations.
\end{abstract}

%% file: 1_introduction.tex
\section{Introduction}
\label{sec-introduction}
Voice recording is an essential information-sharing approach, which is benefiting many aspects of our daily life. Nowadays, smartphones and Internet-of-Things (IoT) devices equipped with microphones allow people to record voice anytime and anywhere. 
However, the growing presence of unauthorized microphones has led to numerous incidences of privacy violations. Off-the-shelf microphones are widely available and can be deployed to steal users' biometric traits (e.g. voiceprints) or private conversations. Thus, 
unauthorized voice recording has become a serious societal issue~\cite{li2020patronus}.

Recent studies~\cite{scramble_noise, li2020patronus} attempt to disrupt unauthorized voice recording by emitting an ultrasonic scrambling noise wave (i.e., a jamming signal) to obfuscate the superposed voice. 
\hanqing{
However, the scrambling noise wave is generated using the low-level acoustic signal features that are irrelevant to the speaker identity. Consequently, other benign microphones in the reception range will also be jammed, most of the time undesirably. 
In fact, the use of such voice jammer in public spaces is prohibited and unlawful (violation of \emph{47 U.S.C. § 333}), since it poses serious risks to critical public safety communication. Moreover, if the attacker learns the frequency pattern of the scrambling noise wave, the attacker can deploy an additional microphone to nullify the noises and record them illegally.}
\hanqing{To allow users to secure their voices lawfully without intervening in others' microphones/recorders usage, we propose} \emph{\ours (Neural Enhanced Cancellation)}, which only jams \emph{a specific target speaker's voice} from the recording of any microphones nearby. 
%




\begin{figure}
    \centering
    \includegraphics[width=0.48\textwidth]{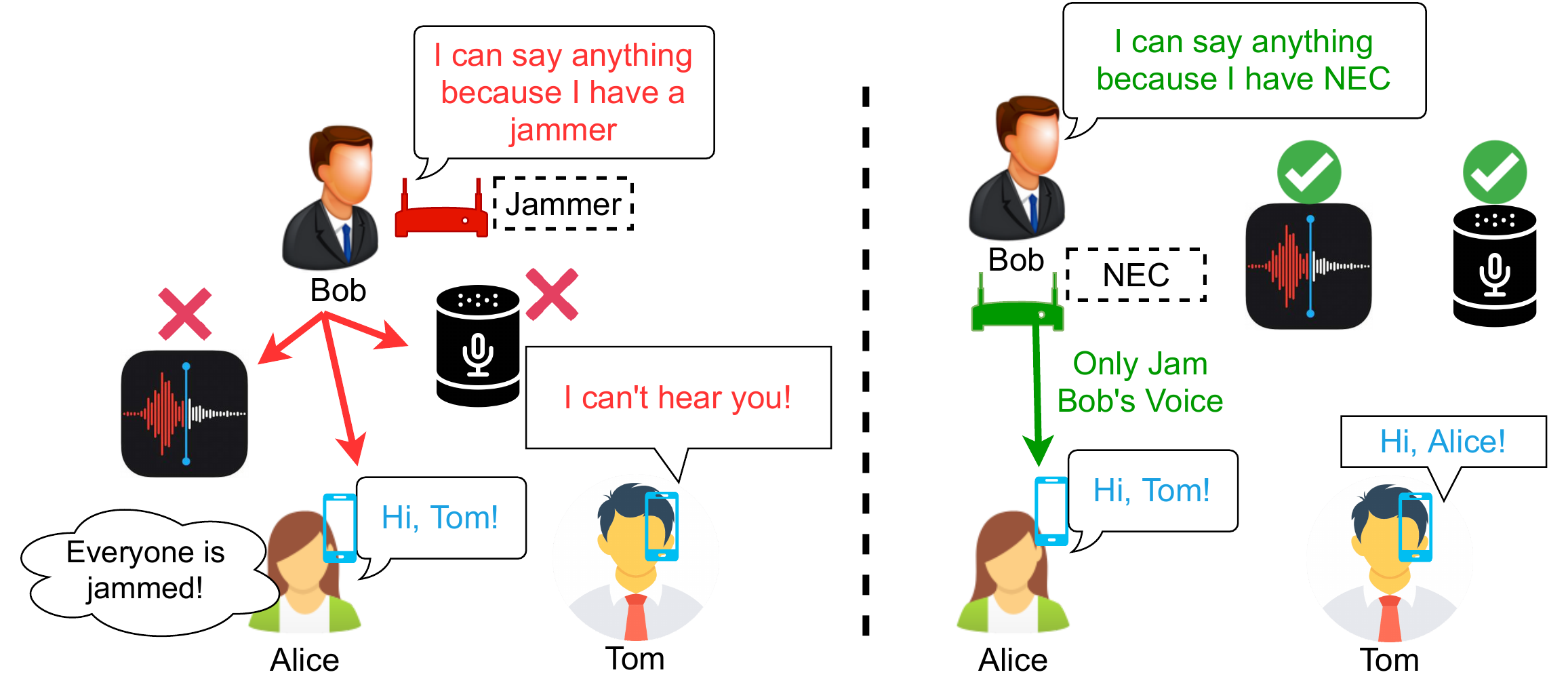}
    \caption{\ours cancels the target speaker's (e.g. Bob's) voice without intervening other communications.}
    \label{fig:cover_1}
    \vspace{-15pt}
\end{figure}

\hanqing{
Figure~\ref{fig:cover_1} illustrates the necessity of deploying \ours instead of the commercial audio jammer. Consider that Bob is initiating a private conversation 
in a public area (e.g., cafe or work office), in order to prevent his speech from being leaked, he turns on a commercial jammer to obfuscate all the surrounding input devices. The left sub-figure shows that during the attack, other applications such as voice reminder, voice assistants, and phone calls are all effectively disabled by Bob's jammer, which is not only unlawful but also annoying to other users. In contrast, if Bob deploys \ours, 
only his speech is imperceptible by the others' microphone, while other users can still safely use their voice applications
as usual.}

\begin{figure}
    \centering
    \includegraphics[width=0.35\textwidth]{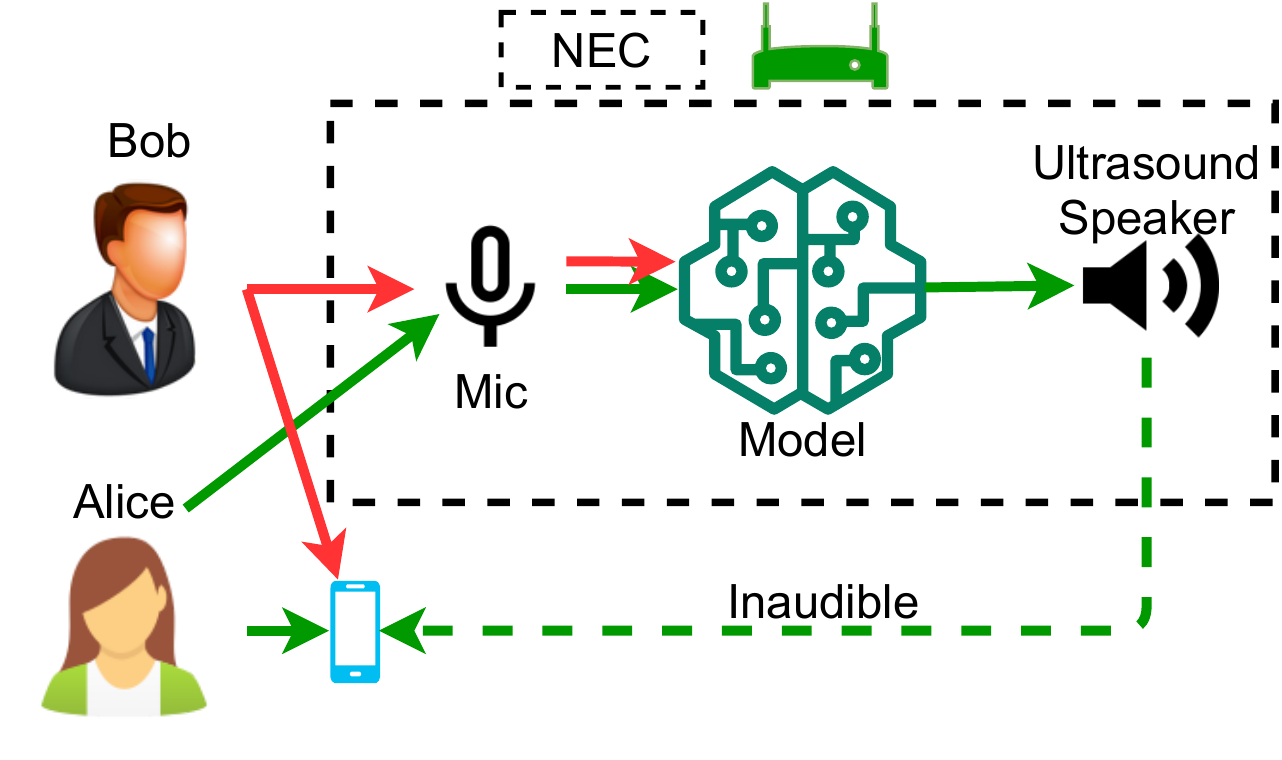}
    \vspace{-15pt}
    \caption{The voice stream flow of \ours}
    \label{fig:cover_2}
    \vspace{-15pt}
\end{figure}

\hanqing{
Generally, \ours is composed of a microphone, a neural network model, and an ultrasonic speaker. Figure~\ref{fig:cover_2} entails the components of \ours, the red lines demonstrate the target speaker's voice (e.g., Bob's voice), while the green lines represent Bob's irrelevant voice (e.g., Alice's voice, background noise, and model processed voice). Our goal is to make Bob's voice unrecognized/unrealized on Alice's phone/recorder. At the very beginning, the microphone perceives both Bob's and Alice's voice. Then, we feed the mixed audio to our proposed deep neural network (DNN) model. Note that, comparing to the existing systems that utilize low-level acoustic signal (such as Gaussian noise or scrambling noise), we use the DNN model to extract the high-level speaker-specific vocal features 
for differentiating Bob's voice from the mixed recordings. The output signal of the DNN model is marked as \emph{shadow sound}, which is then modulated to ultrasonic frequency to make it inaudible to other users. 
Subsequently, Alice's phone will receive a combination of Bob's voice, Alice's voice, and the inaudible shadow signal generated by NEC. The signal combo will yield a mostly undisturbed sound for Alice. 
}

\begin{figure*}
\centering
\includegraphics[width=\textwidth]{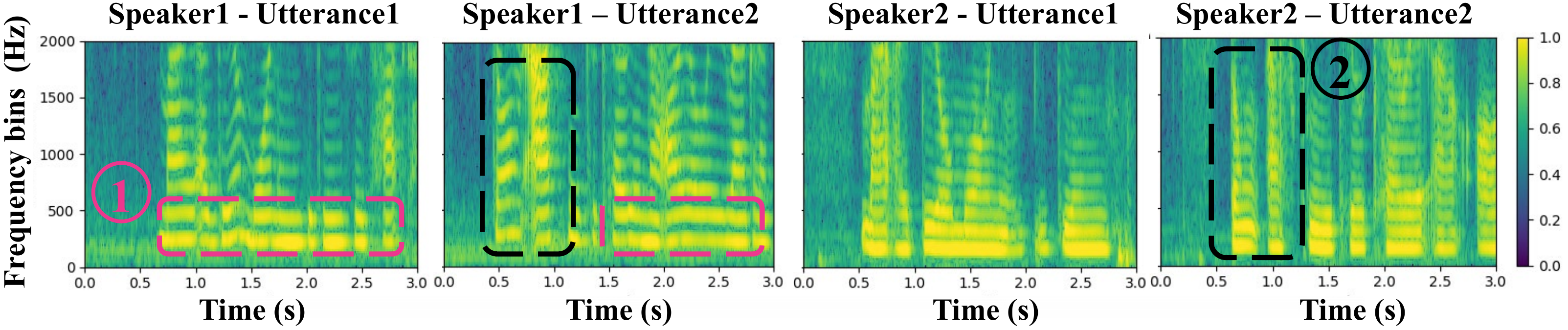}
\vspace{-15pt}
\caption{Distribution of formants across spectrograms, representing the speaker-specific but utterance-independent timber pattern. Utterance1: ``My ideal morning begins with hot coffee.'' Utterance2: ``Don't ask me to carry an oily rag like that.''}
\label{fig:observations}
\vspace{-15pt}
\end{figure*}

We have four main design goals as follows:
\begin{itemize}
    \item \emph{Utterance-independent Vocal Feature Extraction.} For a target speaker, we need to train our DNN model with the speaker's reference audio before the deployment. To alleviate the training overhead across different scenarios, the speaker's vocal features should be independent of his/her utterances. As such, we can deliver a one-fits-all DNN model, which is trained once and easily transferred.
    \item \emph{Microphone-aware End-to-end DNN Training.} The shadow sound is superposed on the speaker's voice at the microphone.
   To make the superposition more effective, 
    we need to design an end-to-end training pipeline that aims to maximize the effectiveness of the superposed shadow sound.
    \item \emph{Low-latency Shadow Sound Generation.} We will modulate a shadow sound onto the ultrasonic frequency to make it inaudible. However, the processing delay may degrade the shadowing efficiency due to the feature mismatch between the speaker's voice and the generated shadow sound. Thus, we need a DNN model that is computationally efficient. 
    \hanqing{
    \item \emph{Synchronization-free.} To cancel Bob's voice on other devices, it typically requires the synchronization of the arrival time of the shadow sound, Bob's sound, and Alice's sound. However, it is challenging to synchronize them
    (without modifying Alice's devices). 
    Therefore, we need a synchronization-free approach for voice cancellation. 
    }
\end{itemize}
To achieve all four goals, we first explore the human vocal principle and observe the speaker-specific but utterance-independent formants of the audio spectrogram from ten speakers using various speech contents. Then, we design a DNN model to generate a shadow sound by imitating the superposition of multiple waves at the microphone. The DNN includes the speaker encoder and selector for feature reference and extraction. Moreover, we analyze the delay bound and compress the DNN layers to guarantee that the processing delay can meet the requirement on various devices (e.g., mobile, Raspberry-Pi).

We implement \ours using commercial off-the-shelf (COTS) ultrasound transceivers and evaluate its performance in different real-world scenarios. 
In the experiment, we run a benchmark testing using 
a public speech corpus dataset and two real-world case studies. The evaluation results demonstrate that \ours effectively mutes the target speaker at a microphone by causing a 200\% word error rate under Google's voice-to-text service without interfering with others’ conversations.
Our contributions are summarized as follows:
\begin{itemize}
    \item 
    \ours is the first practical speaker selective cancellation system, which aims to protect the target speaker's voice without interfering with other microphones in presence. 
    \item We explore the human vocal principles and design a DNN model to imitate the superposition of waves at the microphone, which produces the speaker-specific but utterance-independent shadow audio in real-time.
    \item We implement \ours and extensively evaluate its performance with the benchmark and user studies. The results show its superior performance in comparison with state-of-the-arts systems. The demos can be found on our project website: {\url{https://nec-app.github.io/}}.
\end{itemize}

The paper is organized as follows. $\S$\ref{sec-related-work} summarizes the related work. We then introduce the background and preliminary of our vocal system in $\S$\ref{sec-background}, followed by the system design of \ours in $\S$\ref{sec-system-design}. The implementation and evaluation are further presented in $\S$\ref{sec-implementation} and $\S$\ref{sec-evaluation}. We conclude the paper in $\S$\ref{sec-conclusion}.

%% file: 2_related_work.tex
\vspace{-5pt}
\section{Related Work}
\label{sec-related-work}
\noindent
\textbf{Microphone Jamming:} Microphone jamming has been proposed~\cite{Jamming, scramble_noise, li2020patronus} to protect private conversations. To avoid the recording of private conversations, a pre-configured audio jammer is deployed to emit the scrambling noise waves to disrupt the speech recording. Specifically, Chen et al.~\cite{Jamming} adopt the white noise to distort the microphone recordings, while Tung et al.~\cite{scramble_noise} explore the sound masking with the specially designed scramble noise to obfuscate the spoken sensitive information.  Patronus~\cite{li2020patronus} emits ultrasound to generate the scrambling waves at the recorder without introducing human-sensitive noise.
In contrast, rather than canceling and jamming by the low-level signal features (e.g., frequency, phase), we use high-level human vocal features to generate a shadow sound for speaker-selective jamming. 

\vspace{1mm}
\noindent
\textbf{AI-augmented Speaker Diarization:} AI plays important role of processing signal~\cite{li2020wihf, li2021deep,guo2020deep,zhu2018indoor}.
Recent studies~\cite{wang2018speaker, voiceFilter, Spot_the_conversation_speaker_diarisation_in_the_wild} propose AI-based speaker diarization, a process to partition multi-speaker audio into homogeneous single speaker segments based on the speaker identity. It effectively solves ``who spoke when" in a multi-speaker scenario. 
Several audio embedding models have been proposed for speaker-specific feature extraction, including speaker factor~\cite{speaker_factor}, i-vector~\cite{i_vector_1,i_vector_2}, and d-vector~\cite{wang2018speaker,fully_supervised_speaker_diarization,voiceFilter}. Based on these features, a number of classification models have been designed to extract the speaker-specific embedded audios, such as clustering algorithms~\cite{wang2018speaker,i_vector_1,i_vector_2}, DNN model~\cite{fully_supervised_speaker_diarization,voiceFilter}, and even an integrated model with visual information (e.g., lip movement and face recognition)~\cite{visual_1,visual_2,Spot_the_conversation_speaker_diarisation_in_the_wild}.
However, these methods cannot be adopted in our scenario. First, all existing speaker diarization models are used for post-processing after the audio is recorded, but we need to deal with voice cancellation in an end-to-end fashion.
Additionally, the processing delay is an important factor to guarantee an effective shadow sound generation, which has been ignored by these post-processing models. In this work, 
we design the adaptive features, DNN structures, and training methods to realize an end-to-end voice cancellation system to protect a target speaker's voice.

%% file: 3_0_background.tex
\section{Background of Vocal System}
\label{sec-background}

\noindent
\textbf{Observations:} To illustratively show the harmonic components of a sound induced by the physical vocal system, we first collect four audios from two volunteers. Each volunteer records two audios of their utterances of two sentences: ``\emph{my ideal morning begins with hot coffee}'' and ``\emph{don't ask me to carry an oily rag like that}''. For each audio, we derive the corresponding formants~\cite{formants} via FFT for each frame with a duration of 20 ms. The rationale is that the duration of a typical phoneme is longer than 20 ms, representing the maximal frame length~\cite{caField}. Thus, each frame is dominated by the harmonic components of sustained tones, i.e., the number and relative intensity of the upper harmonics in the sound. 

The results are presented in Figure~\ref{fig:observations}. We can observe the consistent formants of each speaker with various spoken contents. For example, the similarity of the resonant frequency and the relative intensity of formants of different utterances from the same speaker can be observed in area \textcircled{1}, shown in red boxes. Hence, these characteristics are utterance-independent. Conversely, area \textcircled{2} in black boxes implies the distinct distribution of speaker-specific formants, which can also be observed across multiple spectra of various frames.

\vspace{1mm}
\noindent
\textbf{Validation:} Based on the observations above, the remaining challenge is to quantify the utterance-independent but speaker-specific feature in audio spectrograms (i.e., area \textcircled{1} and \textcircled{2}), namely timbre pattern~\cite{timbre}. 
To guarantee the phonetically balanced state in the \rev{timbre},
we first average the dynamic influence of individual phonemes by computing the averaged spectrum for all frames, namely \textbf{Long-time Average Spectrum (LAS)}~\cite{LTAS,caField}. 
 LAS can average out the dynamic characteristics associated with various phonemes such as the motion of the articulators~\cite{speech_principal}. 
Suppose the spoken content for each person is divided into $M$ frames with the duration $T$ in time, the LAS $F(w)_{LAS}$ can be formulated by averaging the spectrum of each frame:
\begin{align}
\label{eq:1}
F(w)_{LAS}=\frac{1}{M}\sum_{m=1}^M\mathcal{F}(f_m(t)), 
\end{align}
where $\mathcal{F}$ denotes the FFT, and $f_{m}(t)$ is the frame waveform signal with the duration $T$. 

\begin{figure}
    \centering
    \includegraphics[width=0.35\textwidth]{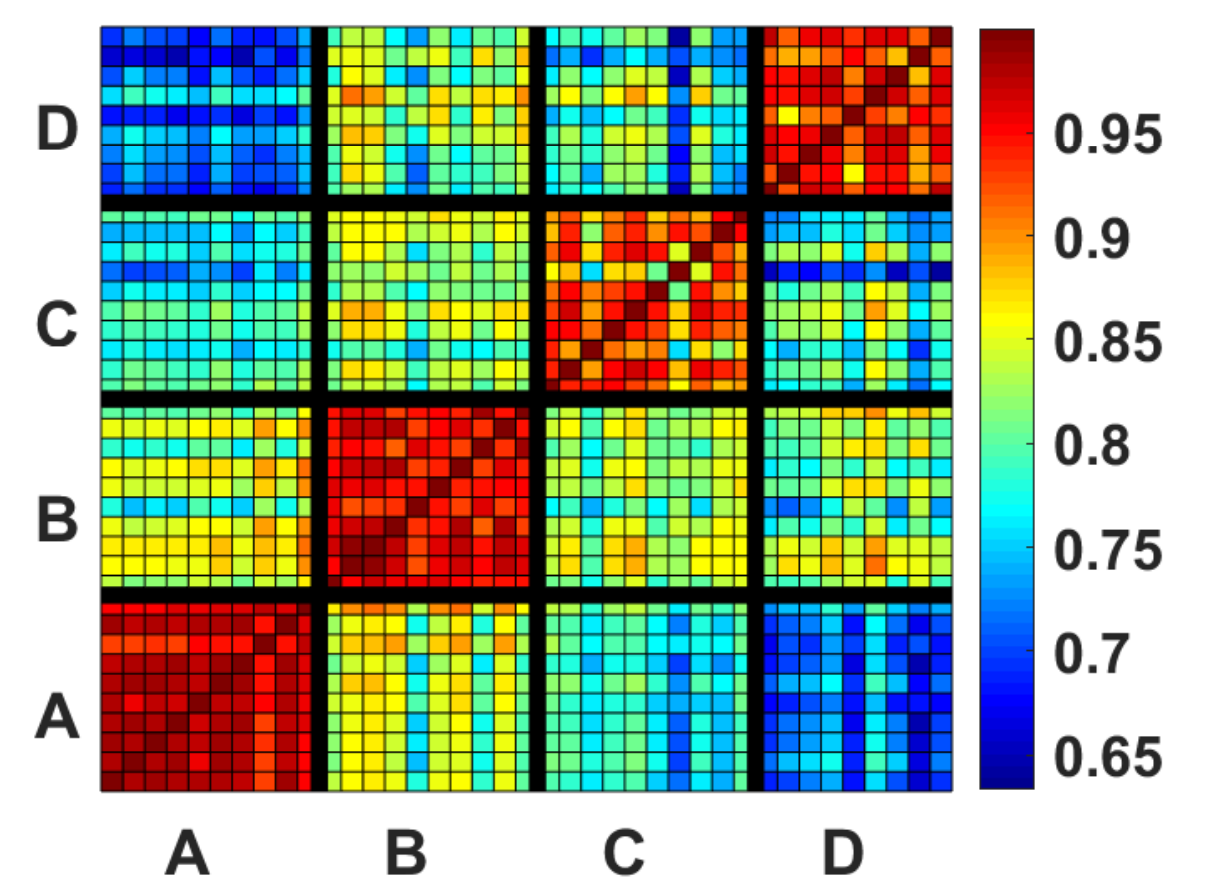}
    \caption{Pearson correlation matrix of the long-time average spectrum of 10 different utterances from 4 speakers. }
    \vspace{-10pt}
    \label{fig:correlation_matrix}
\end{figure}

\hanqing{
To visualize the distinctive LAS features for different speakers, we compute the LAS of four speakers (e.g., A, B, C, D), with every speaker requested to read the same sentence (e.g., ``don't ask me to carry an oily rag like that''). The results in Figure~\ref{fig:LAS} show that 
every speaker's LAS feature is unique even when their speech contents are the same. 
The distinctiveness of LAS features demonstrates the potential of differentiating voices from multiple speakers.
\vspace{-15pt}
\begin{figure}[h]
    \centering
    \includegraphics[width=0.3\textwidth]{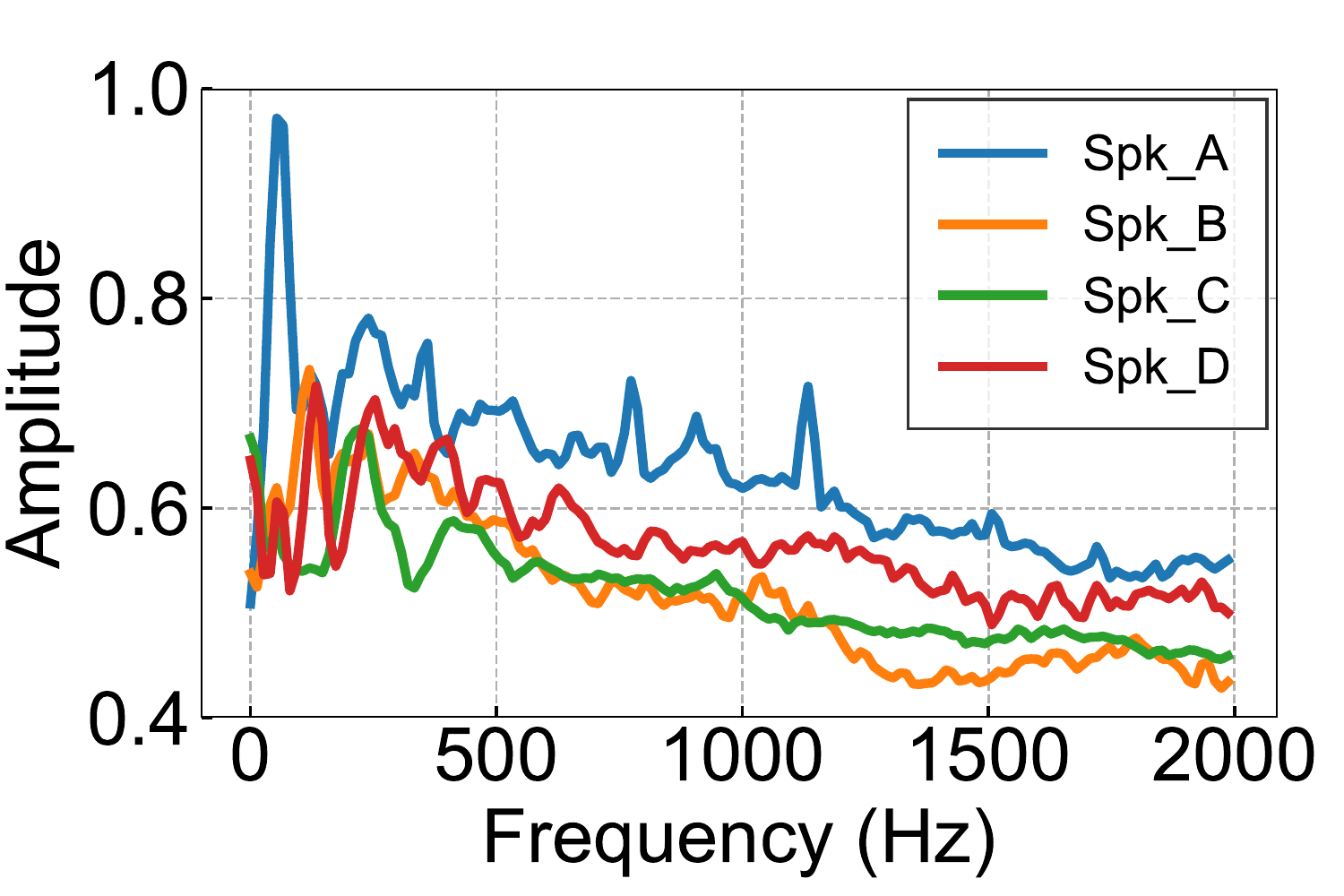}
    \vspace{-5pt}
    \caption{LAS results from four speakers.}
    \vspace{-5pt}
    \label{fig:LAS}
\end{figure}
}
To further verify the utterance-independent but speaker-specific timbre pattern in our computed LAS, we compute the Pearson correlation and deliver the correlation matrix across different speakers and spoken contents. Specifically, we first collect ten different utterances from four speakers (e.g., A, B, C, D) and compute the Pearson correlation across $F(w)_{LAS}$~\cite{caField}. 
As shown in Figure~\ref{fig:correlation_matrix}, the correlation coefficients for the same speaker with different utterances can reach up to 0.96 on average, whereas they are generally below 0.75 across speakers, even with the same utterances. The former implies the consistency of spectrum across various spoken contents for the same speaker, while the latter indicates the distinct timbre patterns of different speakers, which demonstrates the feasibility of using LAS to quantify the timbre patterns from audio spectrograms of different speakers. 

%% file: 3_system_design.tex
\vspace{-5pt}
\section{\ours System Design}\label{sec-system-design}
As shown before, the voice signals from different human speakers present different spectrum features. Meanwhile, for the same speaker, the spectrum features are consistent across different spoken contents. The remaining challenge is to generate a speaker-specific shadow sound from these spectrum features.

 \begin{figure}[h]
 \centering
 \includegraphics[width=0.4\textwidth]{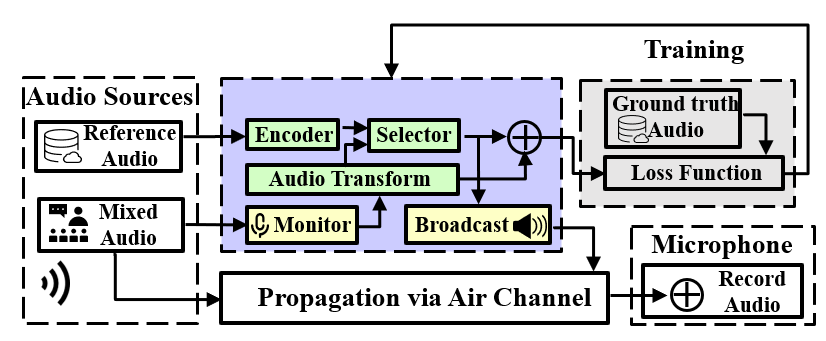}
 \vspace{-10pt}
 \caption{Overview of \ours, which includes the software (green) and hardware (yellow) design as well as the training stage (grey) of our system.}
 \vspace{-10pt}
 \label{fig:overview}
 \end{figure}

\subsection{System Overview}\label{subsec-system-overview}
\noindent
\hanqing{
\textbf{System Pipeline:} The goal of \ours is to cancel \rev{Bob's} voice in the wild (e.g., no one can record Bob's voice in their microphone, and no one is affected by Bob's \ours devices). However, passively canceling Bob's voice on Alice's recorder is very challenging. \rev{Prior work}~\cite{MUTE} takes a great effort to estimate the arrival of Bob's voice through a wireless channel, and compute the inverse signal of Bob's voice before the acoustic signals of Bob arrive. Next, they synchronize Bob's voice with the crafted inverse signal to perform the voice cancellation using rigorous procedures. However, such design relies on the speed difference between wireless signal and acoustic signal. In short-range scenario (e.g., Bob is close to Alice), their work will no longer be effective since the arrival time could be very close. Instead of generating the inverse signal by the prior knowledge of Bob's speech, we propose a \emph{superposition} method to reduce the strength of Bob's sound signals received by Alice's microphone. In other words, \ours produces a shadow signal to be mixed with Bob's voice, which will distill Alice's sound on her microphone.

Figure~\ref{fig:overview} shows an overview of \ours's architecture from audio sources (left block) to the  (Alice's) recording microphone (bottom right block), which serve as inputs and outputs, respectively. The reference audio is the historical recordings of the user, which is prepared to assist the DNN model to separate the voice stream of the user. The mixed audio refers to the audios containing Bob's voice and others' (background) voices. The output of \ours model is a shadow signal transmitted by an ultrasound speaker. The mixed audio and the shadow signal combined together to form the recordings on Alice's microphone. We assume that Alice receives the same mixed audio as the one collected by the \ours's microphone in proximity. 
}

To create a general neural-enhanced framework, 
we first train \ours at the spectrogram level (top right block) in the offline training stage, where a \textbf{$\bigoplus$} operation in the purple block represents the audio spectrogram superposition that combines the outputs of the \textbf{Selector} and \textbf{Audio Transform} modules. \rev{The functionality of \textbf{Selector} is to generate spectrogram that exclude Bob's sound, and the \textbf{Audio Transform} serves to transform the waveforms into spectrograms.}
Then, we convert the shadow spectrogram induced by our selector
into inaudible ultrasound wave via \textbf{Broadcast}. The shadow wave will propagate through the air channel along with the mixed wave ($\S$\ref{subsec-overshadowing-in-the-air}). \textbf{$\bigoplus$} inside the Microphone block indicates the wave superposition of the mixed audio and broadcasting shadow sound at the microphone. Due to the equivalence of audio superposition for wave and spectrogram, the effectiveness of wave superposition is guaranteed for testing scenarios, as mixed audio and shadow sound arrive simultaneously at the microphone. The superposed wave corresponds to the recorded audio which effectively hides the target's (e.g. Bob's) voice. 

\noindent \textbf{Training Stage:} \hanqing{The purpose of model training is to generate \rev{spectrogram that not caused by Bob's voice for any speech context with Bob}. To achieve that, we manually craft mixed audios which contain Bob's voice and other speakers' voice, and use our selector to generate Bob's irrelevant spectrogram.}
To train \ours, we first provide a pre-trained \textbf{Encoder}, which generates the speaker-specific d-vector~\cite{voiceFilter,wang2018speaker} from the reference audio (e.g., 3 audio instances lasting 3 seconds) as reference input for the selector. 
Meanwhile, the mixed audio is processed by the audio transform, which generates a mixed spectrogram as another input of the selector. The rationale of using spectrogram has two folds. First, the \textbf{LAS} feature is effective in distinguishing different speakers based on our previous observation ($\S$\ref{sec-background}); second, the calculation of LAS refers to the procedure of calculating the average spectrum for audio clips, which can be unfolded across multiple clips as a spectrogram. 
We directly feed the mixed spectrogram into our selector, along with the d-vector extracted from the reference input. This can boost the accuracy of DNN in extracting the high-level speaker-specific but utterance-independent vocal features from the mixed sound ($\S$\ref{subsubsec-shadow-audio-generation}). 

\noindent \textbf{Overshadow Stage:}
A key property of \ours is its generalization for deployment in the wild. First, rather than the  cumbersome model-retraining and data collection, only 3 audio instances lasting 3 seconds are required by our one-fits-all model for a new user enrollment. 
Second, due to the linearity of the Fourier Transform ($\S$\ref{subsubsec-system-optimization}), we can transfer the spectrogram superposition into the wave superposition of audios at the microphone to guarantee the overshadowing performance. 
Finally, to avoid the disturbance during the overshadowing of \ours, we further convert the shadow spectrogram into inaudible ultrasound ($\S$\ref{subsubsec-inaudible-shadow-audio-generation}).

\begin{figure}[h]
\centering

\includegraphics[width=0.98\linewidth]{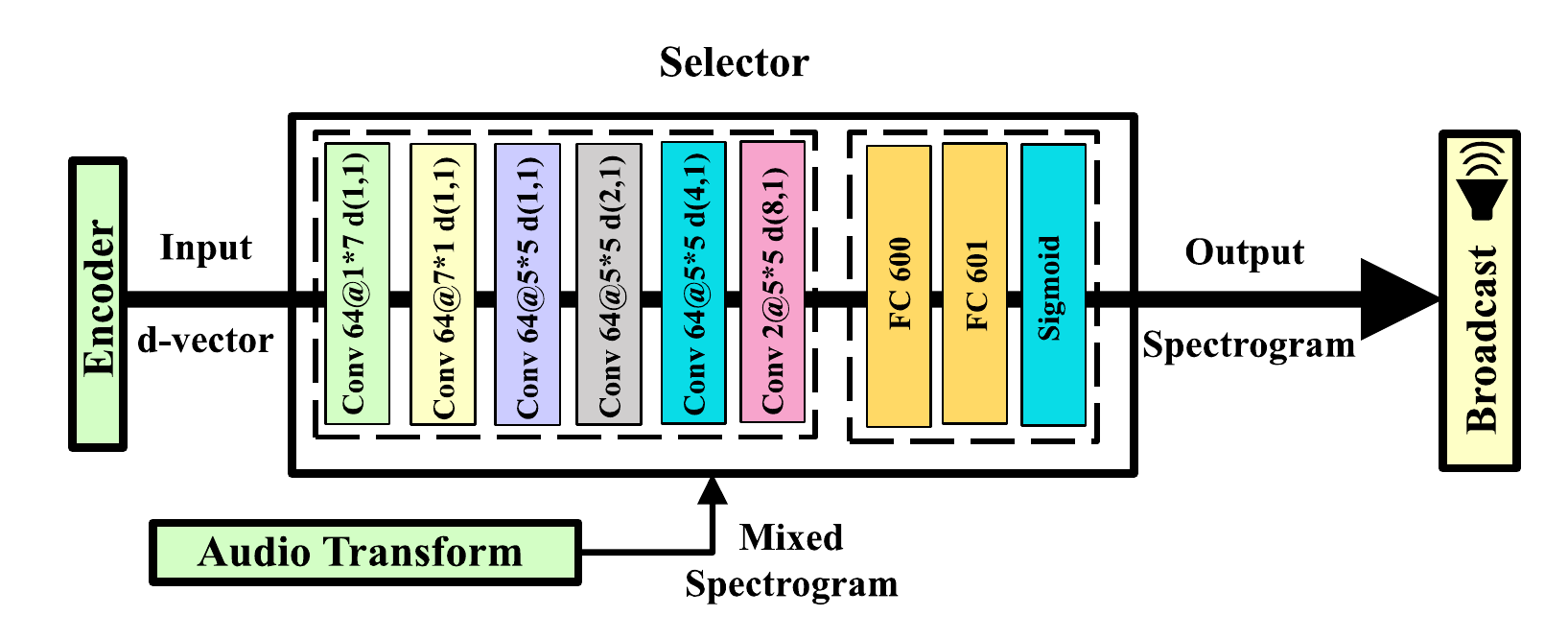}
\vspace{-10pt}
\caption{\ours's DNN Selector generates the utterance-independent but speaker-specific shadow spectrogram by imitating the superposition of waves at the microphone.}
\vspace{-10pt}
\label{fig:flowchart}
\end{figure}

\subsection{Neural Enhanced Selective Speaker Cancellation}
\label{subsec-ai-augmented-speaker-selection}

In this section, we present the design of \ours's DNNs, which aim to utilize the utterance-independent but speaker-specific features to generate the shadow sound. \ours incorporates an efficient \textbf{selector} to produce a shadow spectrogram, and further add Bob's voice through overshadowing onto the mixed spectrogram.

\subsubsection{Architecture of DNN}\hspace*{\fill}
\label{subsubsec-shadow-audio-generation}

\noindent\textbf{Encoder:} 
The encoder module follows the design of \textbf{d-vector} in prior studies~\cite{voiceFilter,10,wang2018speaker}. This module takes the reference audio of a target speaker as input and produces a speaker-specified embedding to allow the \textbf{selector} to filter out the target speaker's voice from the mixed audio spectrogram.

\vspace{1mm}
\noindent
\textbf{Selector:} The purpose of the selector is to produce a shadow spectrogram and further hide Bob's voice by superposing the shadow spectrogram onto a mixed spectrogram. As shown in Figure~\ref{fig:flowchart}, the selector takes d-vector and the mixed spectrogram as input. We formulate the mixture of spectrogram as follows:
\vspace{-10pt}
\begin{align}
    \label{equ-global-optimization}
    S_{mixed}=|\sum_{n=-\infty}^{\infty}x_{mixed}[n]W[n-m]e^{-j\omega n}|,
    \vspace{-5pt}
\end{align}
where the n-sample mixed audio in $\mathbb{C}^n$ is converted into a spectrogram with $t$ $sampling ~points$ and $f$ frequency bins in $\mathbb{R}^{t\times f}$. $W[n-m]$ is the Hann window, and $m$ is the window size. \hanqing{More specifically, the mixed spectrogram is composed of Bob's voice $S_{Bob}$ and background voice $S_{bk}$ (e.g., Alice's voice) as follows:
\vspace{-10pt}
\begin{align}
    S_{mixed}=S_{Bob}+S_{bk}.
\end{align}}
In practice, the input audio lasts 3 seconds with a sampling rate of 16 kHz. The number of samples is 48,000. Also, we set the FFT size as 1,200, resulting in 601 frequency bins. The window length and hop length are 400 and 160, respectively, which generates 299 frames. Then, the shape of $S_{mixed}$ is 601$\times$299, denoted as (F, T), the frequency resolution and frame resolution are 13.31 Hz and 25ms with 15ms overlap. We transpose the mixed spectrogram for further processing and denote the shape of the transposed spectrogram as (T, F). 

With the mixed spectrogram and d-vector in hand, we then utilize them to design a neural network based on our observation in $\S$\ref{sec-background}. Revisiting Fig.~\ref{fig:observations}, the frequency distribution of formants~\cite{formants} and harmonic determine the identity of a given speech (i.e., LAS sufficiently captures the speaker characteristics). \hanqing{Our design goal of the selector is to capture these characteristics with multiple layers of CNNs.}
Prior to building the neural network structure, we propose the requirement for our DNN model as: \textbf{1)} the selector should be able to capture the formants and harmonic feature; \textbf{2)} the selector should consider the consistency of the frequency distribution within the same voice source.

In our DNN design, we only focus on the first three formants since we observed that the lower orders of harmonic have more energy and are more representative for a single speaker. As the bandwidth of the first three formants ranges from 33 Hz to 79 Hz~\cite{fleischer2015formant}, we design the first convolutional layer with 64 filters, whose size is 1$\times$7. The rationale of using this flat filter is to convolve the frequency domain information (F). In particular, each filter covers 93.17 Hz, which is enough to cover the individual formant bandwidth as mentioned previously. Another 64 filters follow, whose size is 7$\times$1, which can cover 115ms (determined by the frame resolution) time-domain feature (T).
It is worth mentioning that the length of phoneme varies from 5 ms to 670 ms based on existing vocal research~\cite{igras2013length}, and the average reading speed for an adult is 184±29 words per minute\cite{trauzettel2012standardized}, i.e., 281$\sim$387ms per word. So the second convolutional layer only serves to explore the detailed information of the phoneme level. 

To further incorporate both F domain and T domain features, we apply a sequence of (5$\times$5) convolutional layers with the dilation ranging from (1,1) to (8,1). The dilation setting on T domain extends the effective range of filters from (5$\times$5) to (5$\times$40), corresponding to 85ms to 610ms. 
This range covers a few words and meets our \textbf{R2} for considering the consistency of frequency distribution. While other studies~\cite{voiceFilter, fully_supervised_speaker_diarization, wang2018speaker} also add extra layers (e.g., LSTM, CNN with larger filter size and dilation shape) for speaker separation task, we consider that those layers play a less important role. For example, a larger filter will introduce irrelevant frequency information and long time span data, when the speaker merely adjusts his/her formants frequency when speaking a single word or a short sentence.

The output of CNNs has the shape of (T, 2$\times$F) since we add a padding layer before the convolutional layer to maintain the shape of feature domain consistency, where 2$\times$F comes from two filters in the last CNN layer. After that, the d-vector is repeatedly concatenated to the output of the last convolutional layer in every time frame. The fused feature embedding will be fed into two fully connected layers. As a result, we get a (T, F) shadow spectrogram. Figure~\ref{fig:flowchart} shows the detailed flowchart of our selector. In total, we only use 6 CNN layers and 2 Fully Connected (FL) layers for the selector model. Compared with the existing models such as \cite{voiceFilter, fully_supervised_speaker_diarization, wang2018speaker}, our model is computationally efficient by eliminating the redundant modules (e.g., LSTM, CNN with larger filter size and dilation shape) unrelated to our research goal.

\subsubsection{Spectrogram-based Overshadowing}\label{subsubsec-system-optimization}
In the overshadowing process, we first feed mixed spectrogram and d-vectors into our selector. Then, we deliver the generated shadow spectrogram to be superposed with the received mixed audio at the microphone. 

\vspace{1mm}
\noindent
\textbf{Shadow Spectrogram Generation:} 
From the point of view of the microphone, the received mixed audio and shadow sound should be superposed to imitate the over-the-air overshadowing at the microphone, formulated as $\textbf{x}_{record}=\textbf{x}_{mixed} + \textbf{x}_{shadow}$. \hanqing{Those vectors represent time-series samples of mixed audio, shadow sound, and recorded audio, respectively.}

Through crafting the shadow sound, our goal is to make the recorded audio as close as background audio (e.g., Alice's sound or environmental noise). 
A straightforward idea is to optimize the shadow sound directly with the audio-level superposition in the time domain. However, there are two drawbacks to the temporal wave superposition. First, the temporal waveform is less representative than a spectrogram.
Second, since the output of our selector is the shadow spectrogram, an Inverse STFT module should be introduced to convert spectrogram to waveform ahead of the loss function, which results in the gradient vanishing issue for back-propagation based on our evaluations. Therefore, we use a shadow spectrogram from our DNN selector for the following overshadowing processing.

\vspace{1mm}
\noindent
\textbf{Superposition for Audio Wave and Spectrogram:} 
The linearity of the Fourier Transform guarantees the equivalence of the temporal wave and spectrogram superposition, which can be denoted as follows:
\vspace{-5pt}
\begin{equation}
\label{eq:linearity}
\mathcal{F}[\sum_{i=1}^na_ix_i(t)]=\sum_{i=1}^na_i X_i(w), \textit{for}\: \mathcal{F}[x_i(t)]=X_i(w),
\end{equation}
where $\mathcal{F}$ denotes the Fourier Transform and $x(t)$ is the temporal waveform signal. Given the linearity of Fourier transform with a coefficient $a_{i}$, we can convert temporal wave superposition into a linear combination of spectrograms as follows:
\vspace{-5pt}
\begin{equation}
    S_{record} = S_{mixed} + S_{shadow}. 
    \label{eq:record_mix}
\end{equation}
The $S_{record}$, $S_{mixed}$ and $S_{shadow}$ correspond to the spectrogram of recorded audio, mixed audio and shadow audio, respectively.
To avoid the gradient propagation issues and expedite the convergence of DNN, the shadow spectrogram from the speaker selector is first normalized before being superposed with the mixed spectrogram. To allow the recorded magnitude to eliminate Bob's voice while retaining other's (e.g., Alice's) voice components, we design the loss function:
\vspace{-5pt}
\begin{equation}
\label{eq:16}
Selector^*_{opt}=\operatorname*{argmin}_{Selector*} ||S_{record}-S_{bk}||_2^2, 
\end{equation}
\hanqing{
where the $Selector*$ denotes the model parameters of our DNN selector, and the $S_{record}$ is the sum of mixed spectrogram and shadow spectrogram. Using the back-propagation with the $L_2$ norm loss, we can derive an optimal parameter $Selector^*_{opt}$ for our DNN selector, which will output an optimal shadow spectrogram $S_{shadow}$. This optimization ensures the resulting $S_{record}$ to be as close to $S_{bk}$ as possible. }

\subsection{Overshadowing Over the Air}
\label{subsec-overshadowing-in-the-air}
\subsubsection{Inaudible Shadow Sound Generation}\label{subsubsec-inaudible-shadow-audio-generation}
Given the shadow spectrogram generated by the trained DNN selector, we can apply the inverse STFT on the shadow spectrogram to derive the shadow sound wave for further broadcasting. To make the shadow sound inaudible for privacy concerns and deployment convenience, we resort to the non-linear property of microphones~\cite{BackDoor,SurfingAttack} to modulate the emitted shadow wave, via the \textbf{Broadcast} module in Figure~\ref{fig:overview}.

\vspace{1mm}
\noindent
\textbf{Non-linearity of Hardware:} The non-linearity property of microphone hardware represents the physical limitations of the diaphragm and the pre-amplifier, which amplify the signals in a non-linear manner.
Mathematically, given an input signal $V_{in}$ to microphone, the output signal $V_{out}$ of the commercial amplifier within the microphone  is not amplified linearly, i.e., $V_{out} \neq A_1V_{in}$, where $A_1$ is the gain for input.
Instead, the output signal is $V_{out}=A_1V_{in}+A_2V^2_{in}+A_3V^3_{in}+\cdots$.
We focus on $A_2V^2_{in}$ of the non-linear $V_{out}$ by ignoring (relatively small) higher-order components~\cite{DolphinAttack,BackDoor}.
 Without loss of generality, let $m(t)$ be a simple tone, e.g., $m(t) = cos(2\pi f_mt)$. We then up-convert the baseband signal $m_{t}$ onto a carrier with central frequency $f_c > 20kHz$. The modulated signal can be written as follows with the power coefficient $\alpha$:
 \vspace{-5pt}
\begin{equation}
   V_{in} = (cos(2\pi f_mt)+\alpha)cos2\pi f_ct. 
\label{equ:v_in} 
\end{equation}

Since $f_c$ is in the inaudible frequency range, the modulated signal $V_{in}$ cannot be heard by humans. 
Given the non-linearity effect, the recorded signal $V_{out}$ will not only contain the linear component $A_1V_{in}$, but also the non-linear component $A_2V_{in}^2$ representing the inaudible but recorded component, denoted as follows:
\begin{align}
    V_{in}^2 & = (cos^2(2\pi f_mt)+\alpha^2 + 2\alpha cos(2\pi f_mt))cos^2(2\pi f_ct)\nonumber\\
          & = \sum_i \lambda_icos(2\pi f_i t)+\mu,
\label{equ:non_linearity}
\end{align}
where $f_i$ denotes frequency components at $f_m$, $2f_c$, $2f_m$,  $2(f_m \pm f_c)$, $f_m \pm 2f_c$ and $\mu$ is a consequent constant.
Given the low-pass filter in the COTS microphone, we can eliminate the high frequency components while retaining the $f_m$ components, where $f_m$ is the baseband frequency of $m(t)$ perceived by a microphone.

\vspace{1mm}
\noindent
\textbf{Shadow Sound Broadcast:}
Then, we can encode our shadow wave $\textbf{x}_{shadow}$ into an inaudible frequency range by modulating it with a carrier whose central frequency is $f_c$.
The broadcast shadow wave can be computed as follows:
\vspace{-5pt}
\begin{equation}
\textbf{b}_{shadow} = \textbf{x}_{shadow}\times cos2\pi f_ct,
\label{equ:broadcast}
\end{equation}
\vspace{-2pt}
\hanqing{
where $\textbf{b}_{shadow}$ refers to the inaudible shadow wave, and $\textbf{x}_{shadow}$ is induced by $S_{shadow}$ from inverse Fourier transform process.} 
More details on modulation setting can be found in $\S$\ref{subsubsec-parameter-study} which evaluates the impact of seven different types of mobile devices.

\subsubsection{Latency Tolerance for Overshadowing}\label{subsubsec-latency-tolerance-for-overshadowing}

\begin{figure}[t]
\centering
\includegraphics[width=0.35\textwidth]{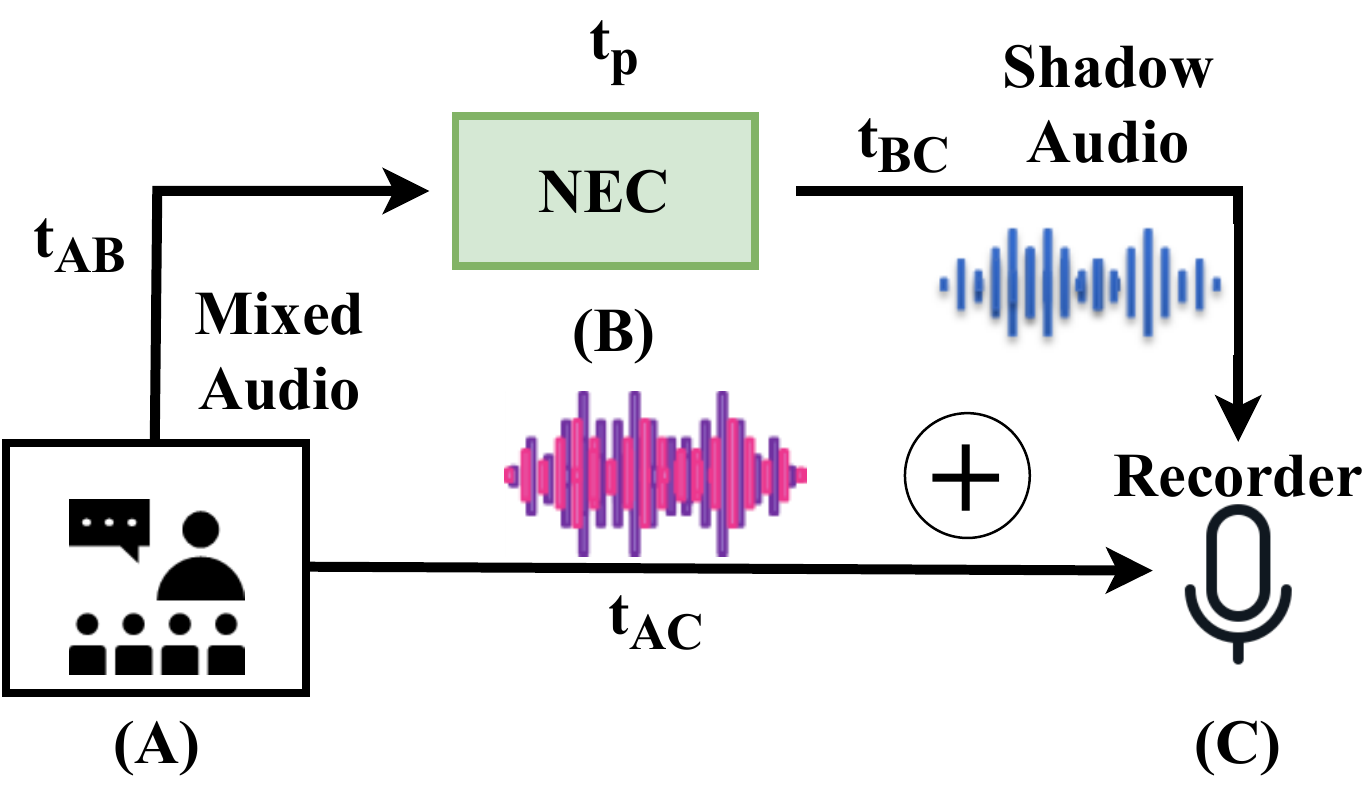}
\vspace{-10pt}
\caption{The illustration of the influence with time offset between the mixed audio and the shadow sound.}
\label{fig:offset_demo}
\vspace{-10pt}
\end{figure}

\noindent
\textbf{Offset Issue in the Wild:} Ideally, the broadcasting shadow wave and the mixed wave arrive at the recorder simultaneously. However, in the real-world scenario, the shadow sound may arrive at the recorder with a slight timing offset due to the propagation delay and system delay, which can be formulated as below:
\vspace{-5pt}
\begin{equation}
    t_{offset} = t_{AB} + t_{p} + t_{BC} - t_{AC},
    \label{equ:time_offset}
\end{equation}
where $t_{offset}$ is composed of the propagation delay $t_{AB} + t_{BC} - t_{AC}$, and system delay $t_{p}$ refers to system process delay, illustrated in Figure~\ref{fig:offset_demo}. Note that the shadow sound received by the recorded is $\textbf{x}_{shadow}$, originating from $\textbf{b}_{shadow}$ based on the non-linearity of hardware ($\S$\ref{subsubsec-inaudible-shadow-audio-generation}).

\begin{figure}[h]
\centering
\subfigure[with time offset.]{\includegraphics[width=0.24\textwidth]{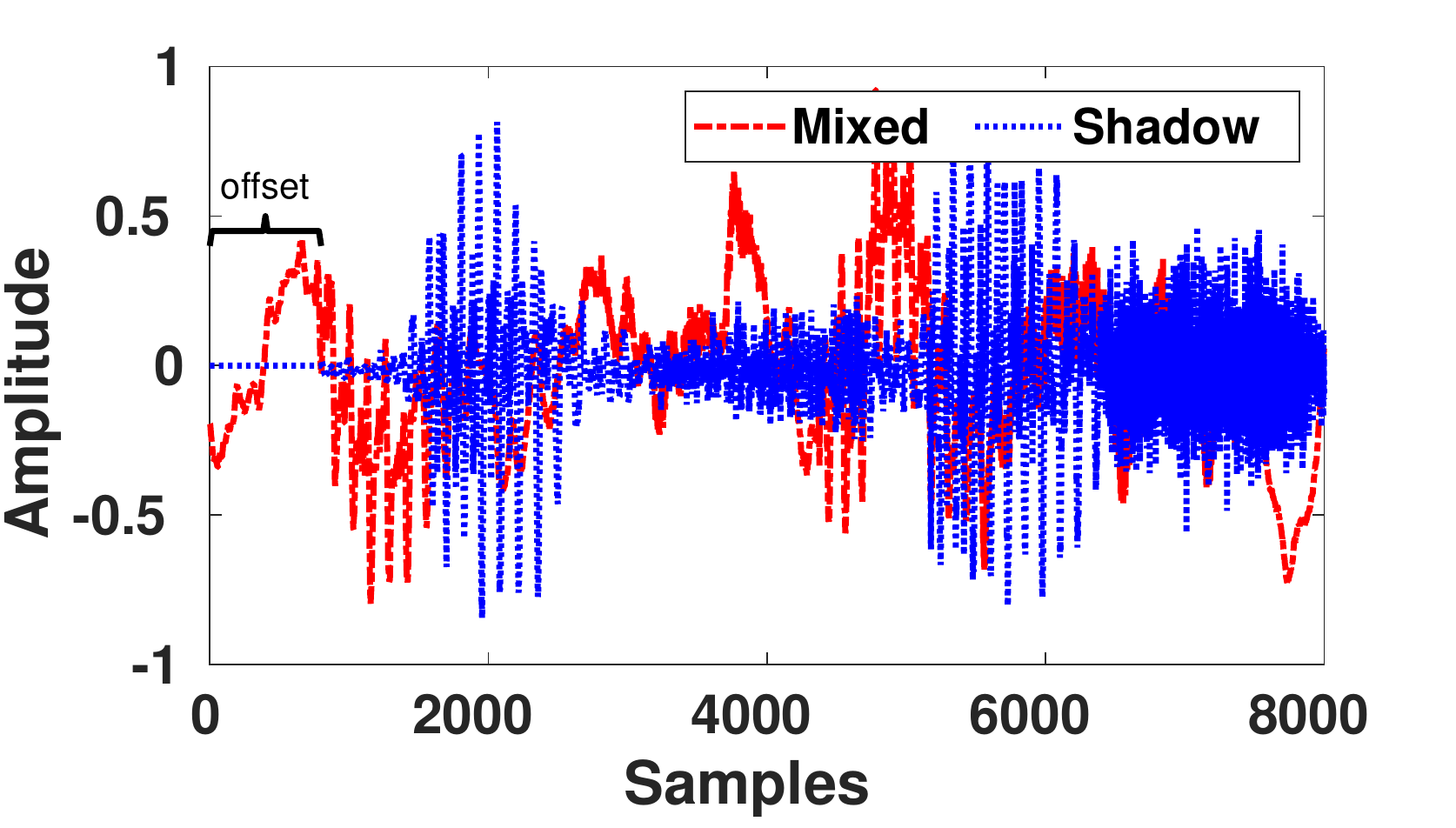}\label{fig:time_offset}}
\subfigure[with power offset.]{\includegraphics[width=0.24\textwidth]{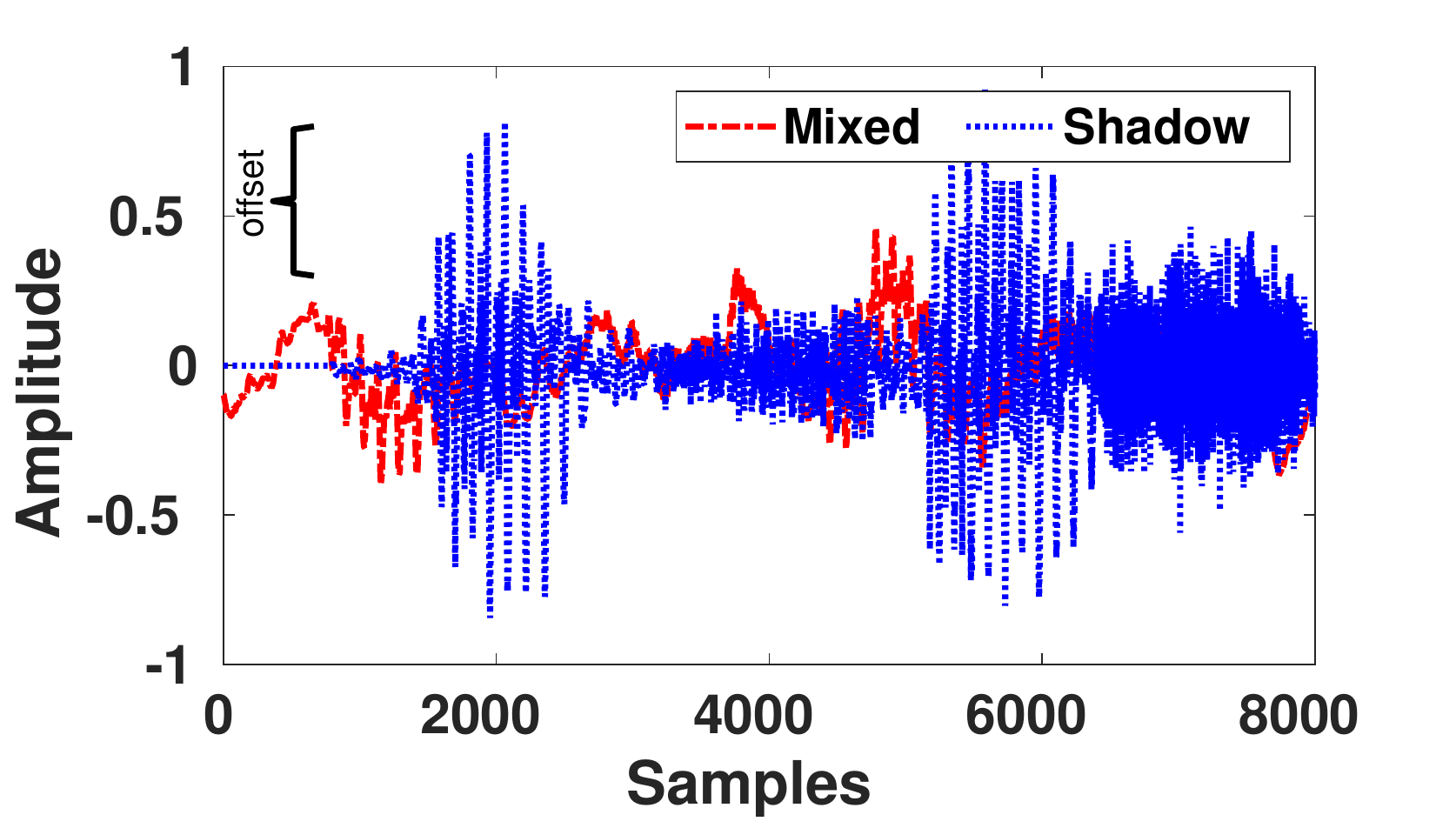}\label{fig:power_offset}}
\subfigure[Cosine Distance]{\includegraphics[width=0.24\textwidth]{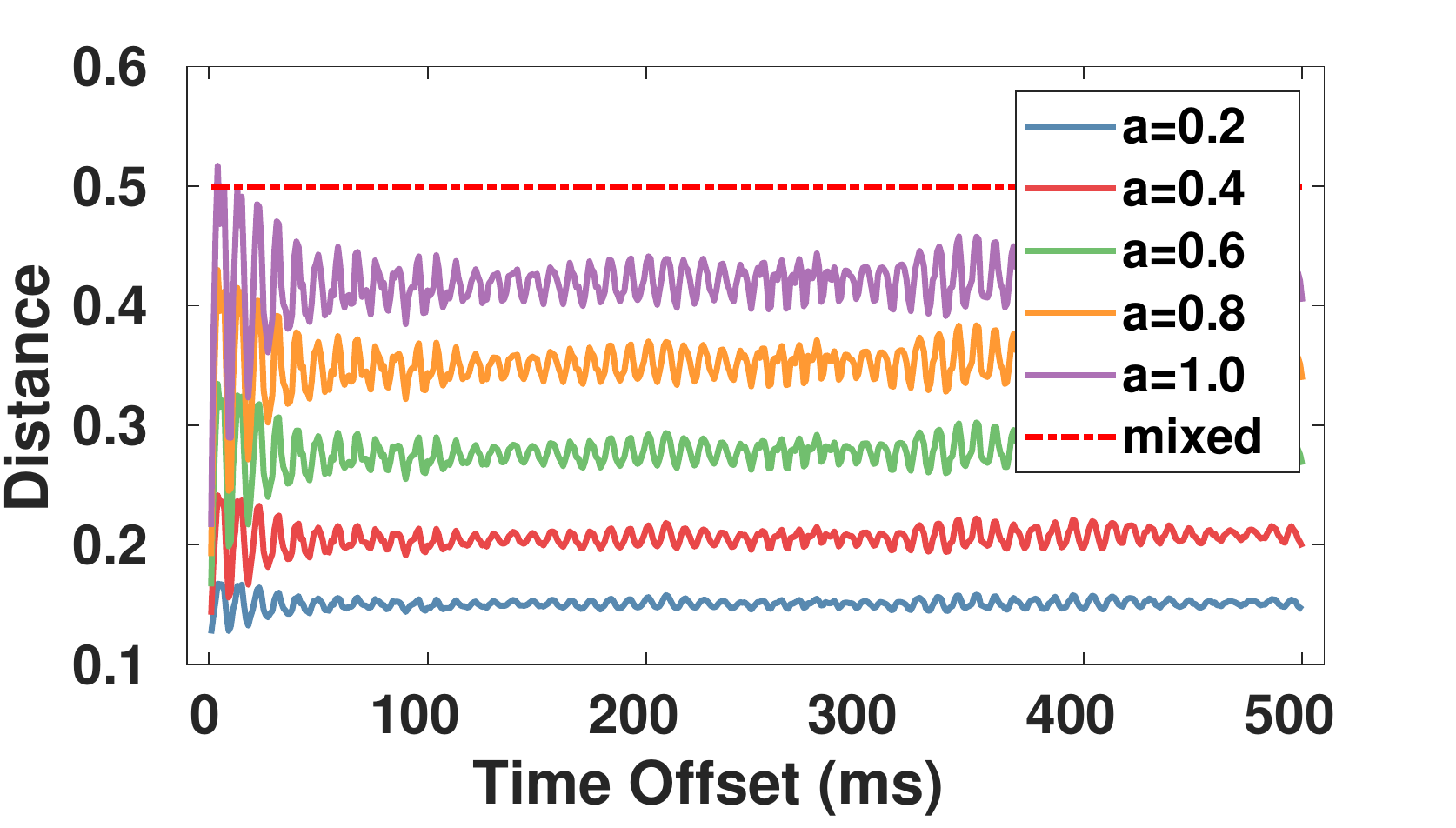}\label{fig:time_offset_dis}}
\subfigure[SDR]{\includegraphics[width=0.24\textwidth]{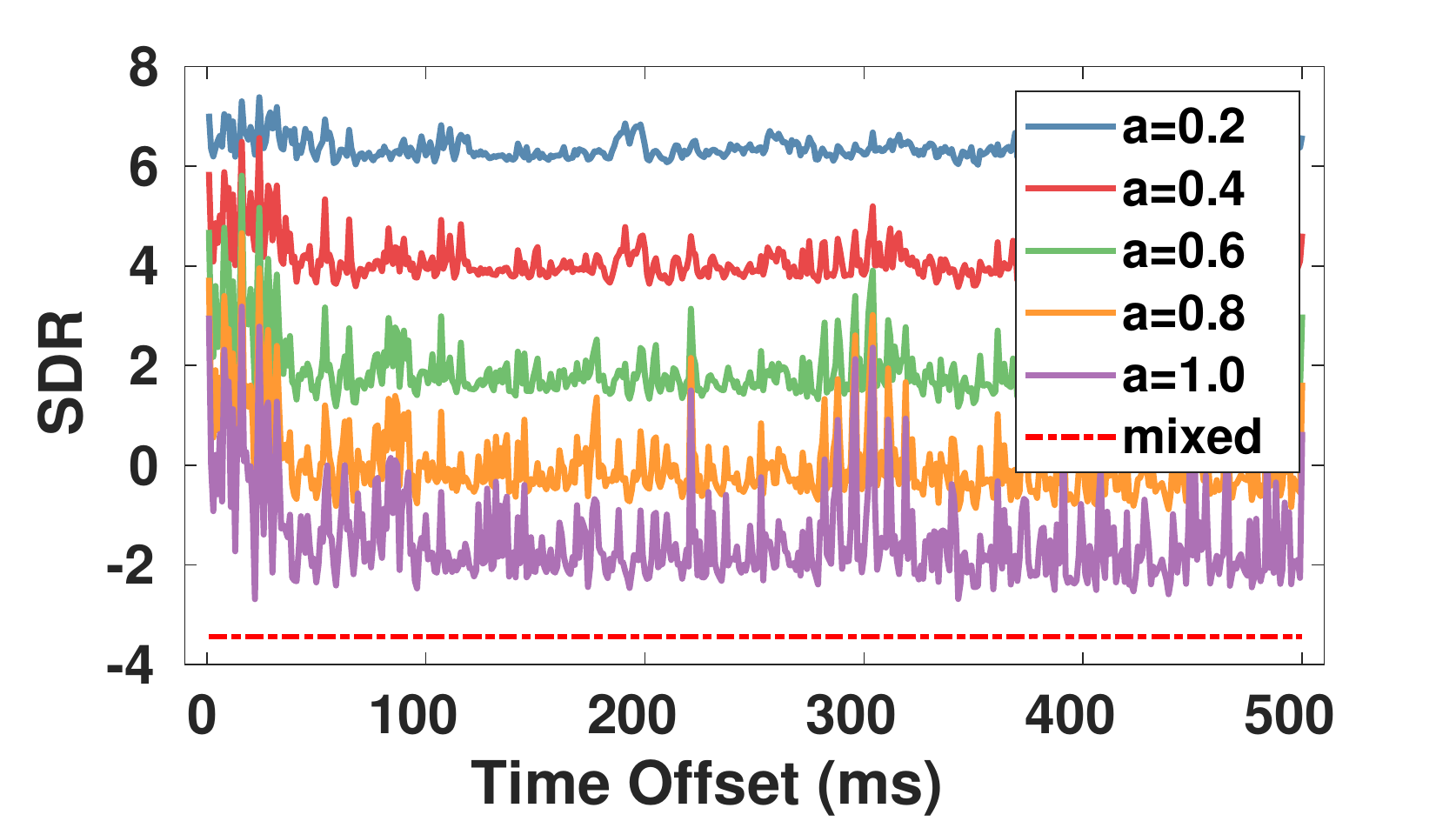}\label{fig:power_offset_sdr}}
\vspace{-10pt}
\caption{Time offsets of overshadowing between the mixed and shadow audios.}
\vspace{-10pt}
\label{fig:offsets}
\end{figure}

Besides the time offset, we still have the power offset between $\textbf{x}_{mixed}$ and $\textbf{x}_{shadow}$, which is introduced by the different attenuations for over-the-air transmission and determined by the original power of mixed audio source and \ours system. To analyze the impact of both offsets, we reformulate the temporal wave superposition to represent the recorded signal as follows:
\vspace{-5pt}
\rev{
\begin{align}
\label{eq:recorded}
x_{record}[n] &= ax_{mixed}[n] +x_{shadow}[n-t_{offset}],\nonumber\\
x_{shadow}[n]&=0 \quad \textit{for} \quad n<0,
\end{align}}
\noindent
where \textbf{$x_{record}[n]$} is the value of $n_{th}$ sample, $a$ is the coefficient to represent power ratio between $x_{mixed}$ and $x_{shadow}$. We set $x_{shadow}$ equal to zero when the shadow sound does not arrive at the recorder side.
Figure~\ref{fig:offsets} demonstrates the time offset and power offset between the mixed signal and shadow signal, respectively. We first collect the mixed signal, which is a 16 kHz recorded audio from a speaker in a noisy car, and extract the first $8,000$ samples. The shadow sound is transmitted through an inaudible frequency carrier and captured by the recorder. Specifically, Figure~\ref{fig:time_offset} illustrates the $800$ samples offset, corresponding to $50$ ms in time delay. We can observe that the mixed audio will be superposed by the shadow wave if $a$ is $0.5$, as shown in Figure~\ref{fig:power_offset}.

\vspace{1mm}
\noindent
\textbf{Tolerance Analysis:} Theoretically, we expect our time offset to be within the one-word duration, such that the offset will not be perceived by the user. Since the average reading speed for an adult is $184\pm29$ word per minute~\cite{trauzettel2012standardized}, i.e., $281 \sim 387$ms per word, our theoretical analysis shows that if the real-time offset is limited within around $300$ms, the recorded audio would be clear to humans. 
For the power offset, a recent study~\cite{yang2019hiding} shows that performing wireless overshadowing attack
requires a power difference between overshadow signals and legitimate ones to be as small as $3$ dB, i.e., $a=0.5$.

\noindent
\textbf{Quantitative Analysis:}
We also measure the cosine distance and Source to Distortion Ratio (SDR)~\cite{SDR,voiceFilter} along with different settings. Figure~\ref{fig:time_offset_dis} shows the cosine distance between the recorded signal at different time offsets, and different power coefficients $a$ of the background (e.g., Alice's sound) at the same time offset. 
\hanqing{For comparison, we also compute the cosine distance for mixed signal and background audio. The shorter cosine distance indicates a greater similarity between record audio and background audio. We can observe that: first, the lower power coefficient leads to shorter cosine distance; second, the time offset within $500$ms does not affect the cosine distance significantly; 
third, if we emit higher power shadow audio than the power of mixed audio (e.g., $a < 0.6$), then it is guaranteed that the recorded audio has high cosine similarity with the background sound; fourth, by applying our shadow audio,  the similarity between record audio and background audio increases. 
The mixed audio without the addition of shadow audio has the largest cosine distance.}

Similarly, Figure~\ref{fig:power_offset_sdr} calculates the SDR between different record signals (with different power offset) and the background audio. The higher SDR indicates less distortion of the audio. 
A reference result is generated by the SDR of mixed signal and background audio. 
This result not only supports our previous observations in Figure~\ref{fig:time_offset_dis} that the lower power coefficient is better but also reveals that the smaller time offset (within $50$ms) results in higher SDR. 

Based on our theoretical and quantitative analysis, we identify the requirements for successfully implementing \ours in a real-world scenario. That is, \emph{the time offset introduced by propagation delay and system latency should be limited to $300$ms. 
To superpose the shadow signal on the mixed signal, the power coefficient is expected to be lower than $0.6$, in which case the desired record audio will be perceived by the recorder.} We further justify the requirements 
in $\S$\ref{sec-evaluation}.

%% file: 4_evaluation.tex
\vspace{-5pt}
\section{Implementation}\label{sec-implementation}
\begin{figure}[!t]
\centering
\includegraphics[width=0.6\linewidth]{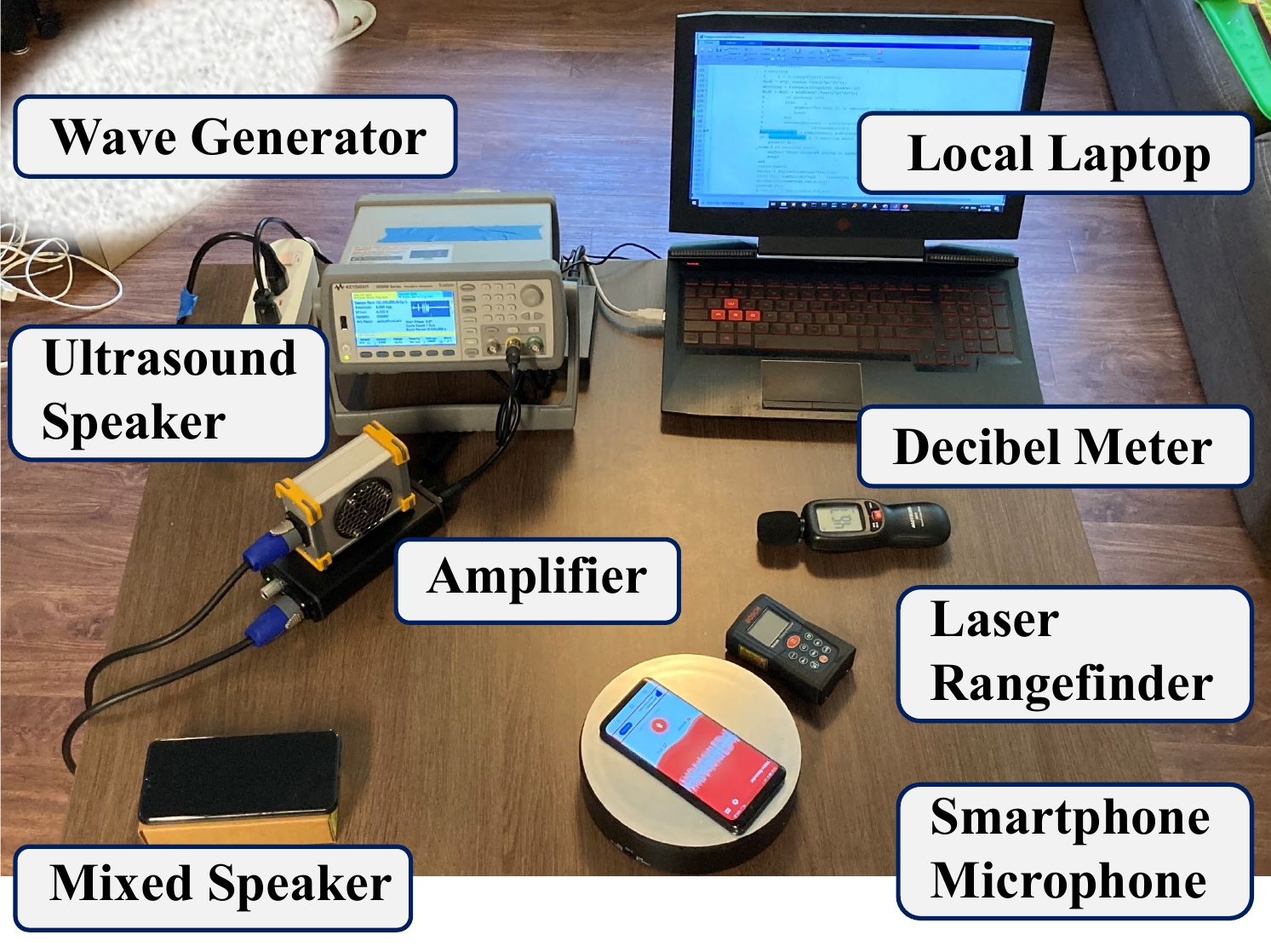}
\vspace{-10pt}
\caption{Implementation and experimental settings.}
\label{fig:setup}
\vspace{-15pt}
\end{figure}

\noindent
\textbf{Experimental Setup:} Figure~\ref{fig:setup} presents the implementation and experimental settings of \ours. In \ours, the input mixed audio is first collected and processed by our trained encoder and selector DNNs, which produces the corresponding shadow spectrogram in Figure~\ref{fig:flowchart}. Then, we transform it into audios and up-convert it into the ultrasound carrier frequency, making it inaudible during broadcasting ($\S$\ref{subsec-system-overview}). We run \ours on a local laptop ($\S$\ref{subsubsec-running-time-analysis}) to generate the shadow spectrogram, which is sent to a Keysight 33500B waveform generator, followed by an ultrasonic power amplifier~\cite{Amplifier} to amplify the inaudible shadow wave. Being transmitted through the air by a wide-band dynamic ultrasonic speaker, Vifa~\cite{Vifa}, the shadow wave is superposed with the mixed audio at a COTS smartphone's microphone. We use a loudspeaker to play mixed audios, i.e., the "Mixed Speaker" in Figure~\ref{fig:setup} emulates a mixed conversation from Alice, Bob, and others. The target's voice will be effectively muted in the final recorded audio. 

\begin{figure*}[!t]
\centering
\includegraphics[width=0.98\linewidth]{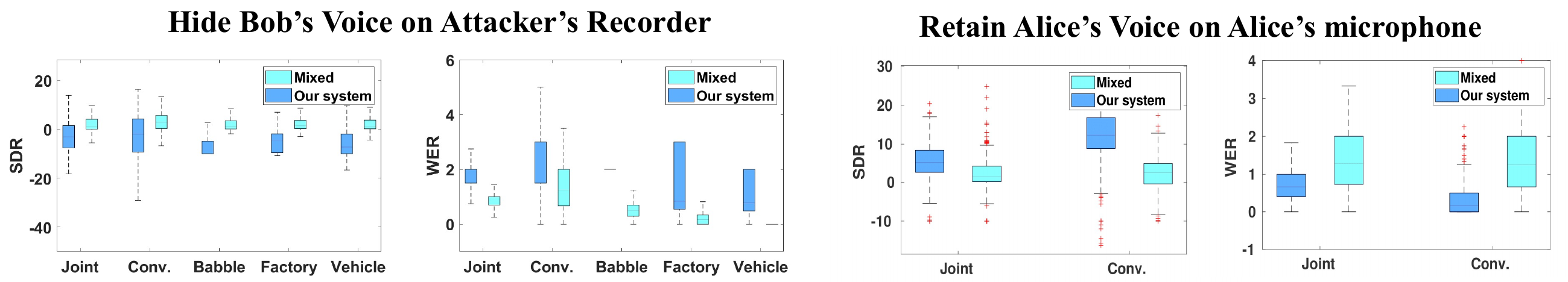}
\vspace{-10pt}
\caption{Overall system performance of our system on three setups across multiple sources of noises.}
\vspace{-15pt}
\label{fig:benchmark}
\end{figure*}

\vspace{1mm}
\noindent   
\textbf{Dataset Compilation:}
Table~\ref{tbl:dataset} summarizes our testing dataset. First, we conduct the \textbf{System Benchmark} by testing the target speaker with the public datasets in controlled environments to verify whether our target speaker's voice can be hidden in the presence of real-world noises. Then, we deploy our system in the wild for a real attack scenario: the target volunteer wants to avoid being recorded while talking in public scenarios, but the COTS microphone can record others' voices normally.

\begin{figure}[h]
\centering
\includegraphics[width=0.6\linewidth]{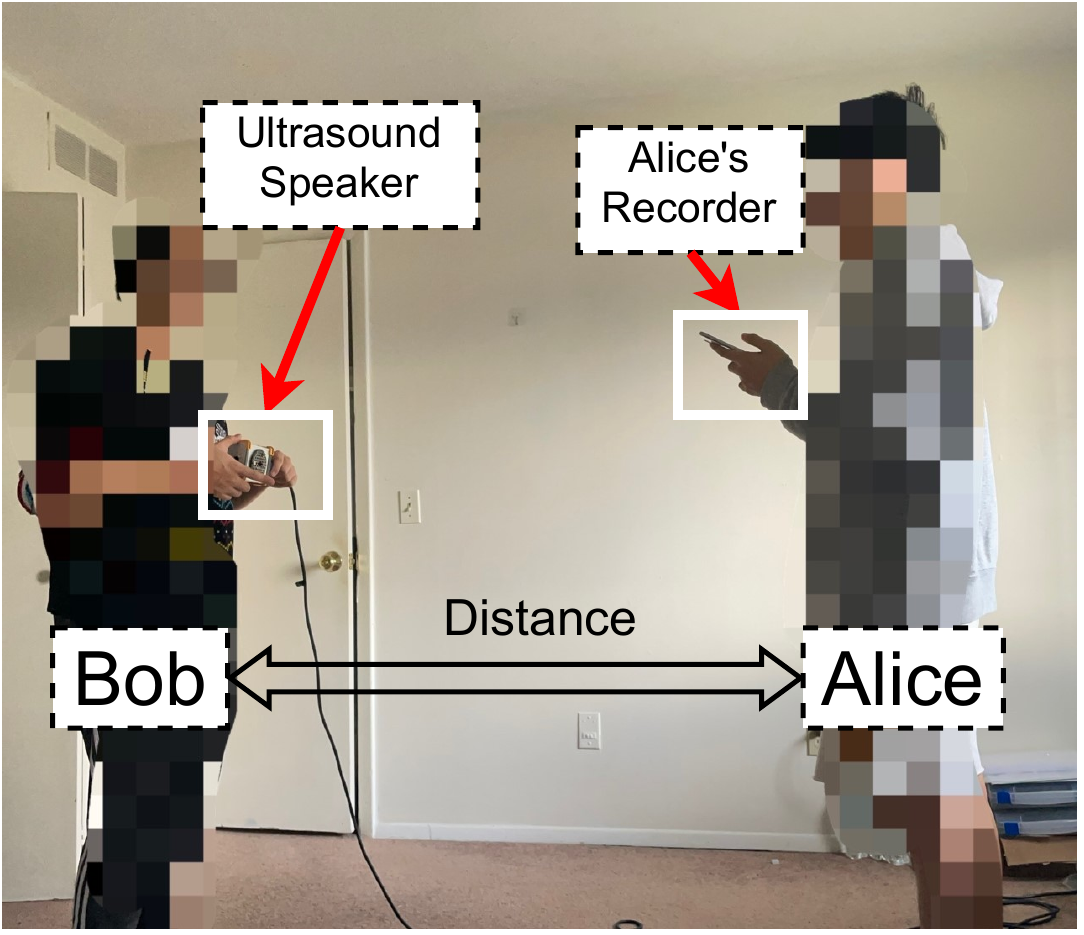}
\vspace{-5pt}
\caption{Hiding Bob's voice from Alice's recording in a real-world scenario. }
\label{fig:realworld}
\vspace{-10pt}
\end{figure}

\begin{itemize}[leftmargin=*]
\item \textbf{Model Training:} Prior to the evaluation of \ours, we train a one-fits-all DNN model for all the defensive scenarios in public. The training dataset is constructed by mixing audios of two different speakers from LibriSpeech~\cite{panayotov2015librispeech}, and mixing target speaker audios with different noises from NOISEX-92~\cite{varga1993assessment}. We provide the background audio that excludes the target speaker and train our model to hide the target speaker's voice, given the mixed audio and reference audios of the target speaker.

\hanqing{
\item \textbf{System Benchmark:} Using the public dataset LibriSpeech~\cite{panayotov2015librispeech} as the corpus source, we first select \textbf{10 target speakers}, we collect 3 audios for each target speaker as their reference audio, and the rest audios of the speaker are treated as normal speech in a real scenario (e.g., Bob's speech). To measure the robustness of \ours, we simulate 
different environments with different types of noises. In order to cover different frequencies of noises (e.g., high-frequency speech and low-frequency ambient noises), the noises from 5 application scenarios are then mixed with the target speakers' voices, which results in 3,190 mixed audios in total. Then, we randomly mix the 10 target speakers' voices with the ones from the other 40 speakers, which generates 560 total instances for the joint conversation.
\item \textbf{User Case Studies-1:} We further collect the user study dataset from \textbf{10 target volunteers}, covering 3 females and 7 males. All volunteers are required to speak 25 sentences, respectively. Analogous to the dataset for the benchmark, we select the reference and test audios randomly, then mix test audios with 4 sources of noise. In total, 160 mixed audios are produced. Then, we randomly mix the audios of 10 target volunteers with the ones from another 18 volunteers to derive the joint conversation dataset.
\item \textbf{User Case Studies-2:} We conduct another user case study to justify the feasibility of \ours in the real world. As shown in Figure~\ref{fig:realworld},
Bob carries the \ours device to hide his sound in the wild. We ask Bob and Alice to speak normally, with volume at $77dB_{SPL}$ from our decibel meter  placed at $5cm$ away from their lips. Then, we record the loudness, SONR, and the proportion of Bob's sound on Alice's recorder (a Moto Z4 phone) at different distances for different cases (with or without \ours).}
\end{itemize}

Note that our testing dataset is disjoint from the training one and reference audios. Thus, the two trained models can be deployed directly with only three arbitrary reference audios from the new target speaker volunteers, avoiding the cumbersome deployment costs (e.g., model re-training and data re-collection)~\cite{Widar3.0,CrossSense}.

\begin{table}[t]
\caption{Testing dataset for benchmark and user cases}
\vspace{-10pt}
\begin{center}
\begin{tabular}{c|cccc}
\hline
Scenario & Source & Freq. & Type & Instance\\
\hline
Joint$^{\mathrm{a}}$ & LibriSpeech & 0-8k & 40/18 & 560/-\\
Conv. & / Volunteers & 0-8k & 40/- &  560/40\\
\hline
Babble$^{\mathrm{b}}$ & \multirow{3}{*}{NOISEX-92} & 0-4k & - & 690 / 40\\
Factory$^{\mathrm{c}}$ & & 0-2k & - & 690/40 \\
Vehicle$^{\mathrm{d}}$ & & 0-500 & - & 690/40\\
\hline
\multicolumn{5}{l}{$^{\mathrm{a}}$Two speakers talk jointly. $^{\mathrm{b}}$100 people whispering.}\\
\multicolumn{5}{l}{$^{\mathrm{c}}$a production hall. $^{\mathrm{d}}$a vehicle running at 120 km/h.}\\
\end{tabular}
\label{tbl:dataset}
\end{center}
\vspace{-15pt}
\end{table}

\vspace{1mm}
\noindent
\textbf{Quantitative Metrics:} To measure the quality of \ours, we consider four main metrics:
\begin{itemize}[leftmargin=*]
    \item \textbf{Source to Distortion Ratio (SDR)}~\cite{SDR,voiceFilter} measures the ratio of energy (in dB) between the energy of the target signal and the errors (induced by the interfering speakers and artifacts) in the mixed signal. It should be low for Bob's voice and high for Alice's voice.
    \item \textbf{Word Error Rate (WER)} is adopted broadly to evaluate the machine translation systems~\cite{WER}. We compute the WER by employing Google's speech-to-text service to transform the acoustic signals into texts. \ours aims to enlarge the WER for the target speaker and minimize it for other speakers (e.g., Alice).
    \item \textbf{User Rating Score (URS)} is the rating for recordings, in which \textbf{10 reviewers} rank the raw mixed and recorded audios of \ours with score 1-5, along with Bob's clean voice as the ground truth. \hanqing{ Specifically, score 5 denotes the best performance, in which reviewers cannot recognize any words of the target speaker (e.g., Bob)}.
    \hanqing{
    \item \textbf{Sound Noise Ratio (SONR)} is used to evaluate the proportion of Bob's sound in the recorded sound. We regard the mixed audio as useful sound and treat Bob's voice as noise. By computing the power ratio between the mixed audio and Bob's sound at different distances, we validate the efficacy.}
\end{itemize}

\vspace{-5pt}
\section{Evaluation}
\label{sec-evaluation}

In this section, we comprehensively evaluate \ours in different   environments with different settings and devices.
\subsection{Overall Performance}
\noindent
\textbf{System Benchmark:} We first evaluate \ours on the public dataset and provide SDR and WER across multiple scenarios in Figure~\ref{fig:benchmark}. When the target speaker, i.e., Bob, expects his voice to be hidden in the recordings,
the recorded audios achieve a lower SDR and higher WER compared with the mixed audios. This shows that our shadow audios can hide Bob's voice reliably, making it unrecognizable by the Google service. Specifically, the median WER increases from 0.894 to 1.798, while the SDR reaches -4.918 dB from 0.997 dB. Note that the WER of the mixed audio is too high to be recognized by the Google service due to the background speeches from other people. Yet, it can still be recognized by humans. Conversely, \ours achieves a higher WER by hiding Bob's voice using the shadow wave, making it even unrecognizable for humans. We further verify its efficacy in the user studies below. 

\begin{figure}[h]
\centering
\includegraphics[width=\linewidth]{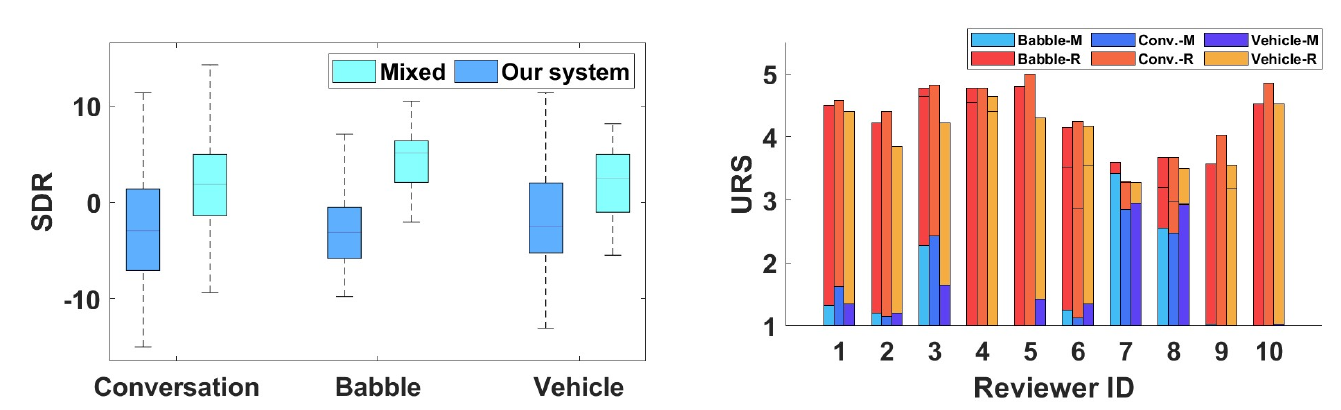}
\vspace{-10pt}
\caption{User study results.}
\label{fig:usercase}
\vspace{-15pt}
\end{figure}

Also, we evaluate the effectiveness of \ours to retain others' voice (e.g., Alice) in Figure~\ref{fig:benchmark}(right). We set the ground truth as Alice's clear voice, and calculate the SDR and WER for the recorded audio and the ground truth audio. The result shows that, compared to the mixed audio which contains Bob's voice, 
we can achieve higher SDR and lower WER for capturing Alice's sound when Bob deploys \ours. 
\noindent
\textbf{User Case Study-1:} Figure~\ref{fig:usercase} shows the performance of SDR and URS for hiding target volunteers' voices in the wild.
We observe a consistent declination in SDR of the recorded audios compared with raw mixed ones. We can hardly recognize the target volunteer's voice in the recorded audios, as the median SDR reaches -4.374 dB, much lower than the SDR of mixed audios at 2.798 dB.
To evaluate the recorded audios comprehensively, we ask 10 reviewers to score the recorded audios and the mixed ones with ranking scores from 1-5. We expect a higher score when fewer utterances of the target volunteer can be recognized. It shows that the average score of the recorded audios can reach $4.034$ for different reviewers. All 10 reviewers give 4 for most of the  recorded audios, while most scores of 1 are given to the mixed audios, except the reviewers 7 and 8. 
\begin{figure*}[t!]
\centering     
\subfigure[Distance = 0.5m
]{\label{fig:a}\includegraphics[width=42mm]{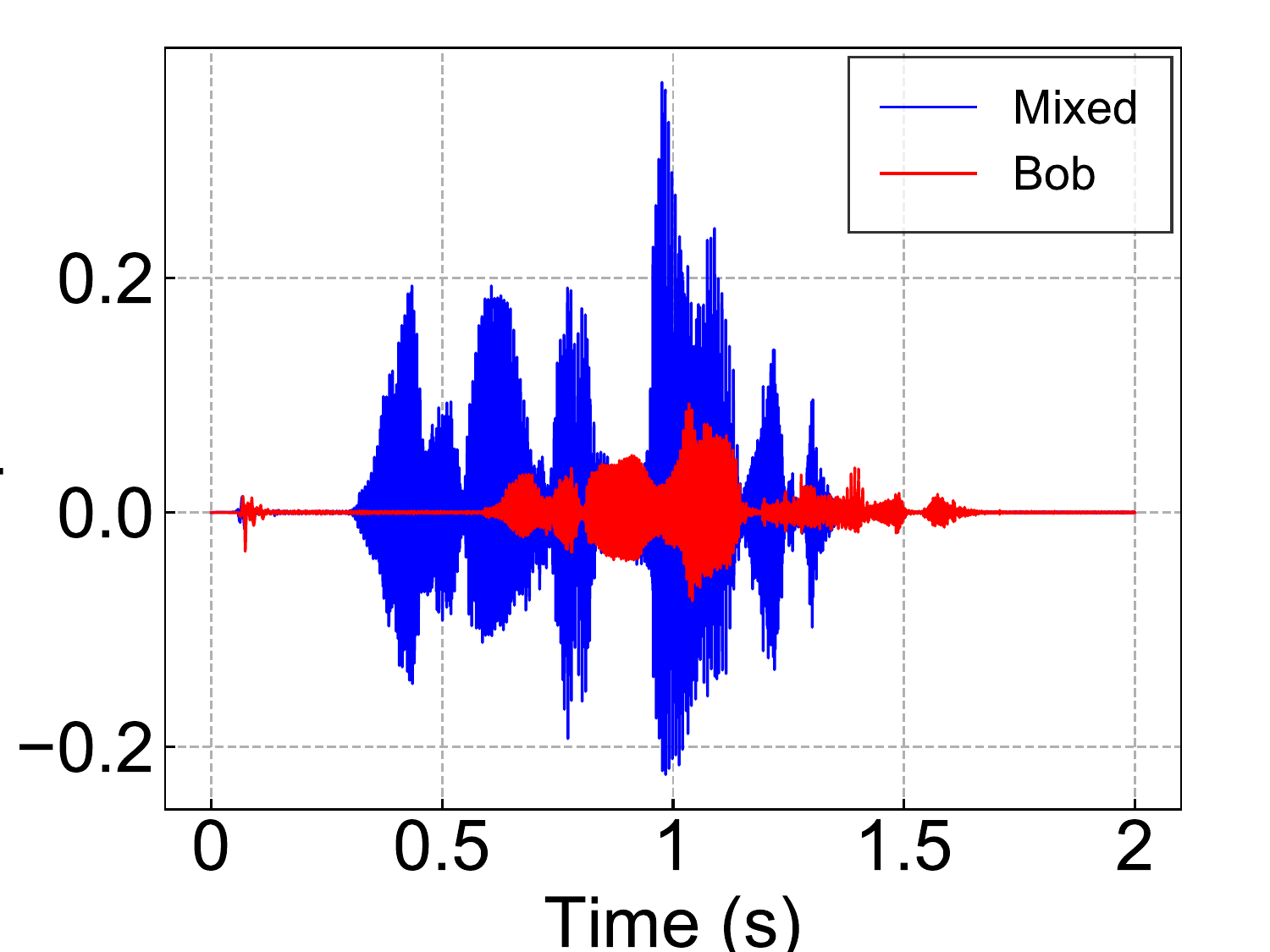}}
\subfigure[Distance = 1m]{\label{fig:b}\includegraphics[width=42mm]{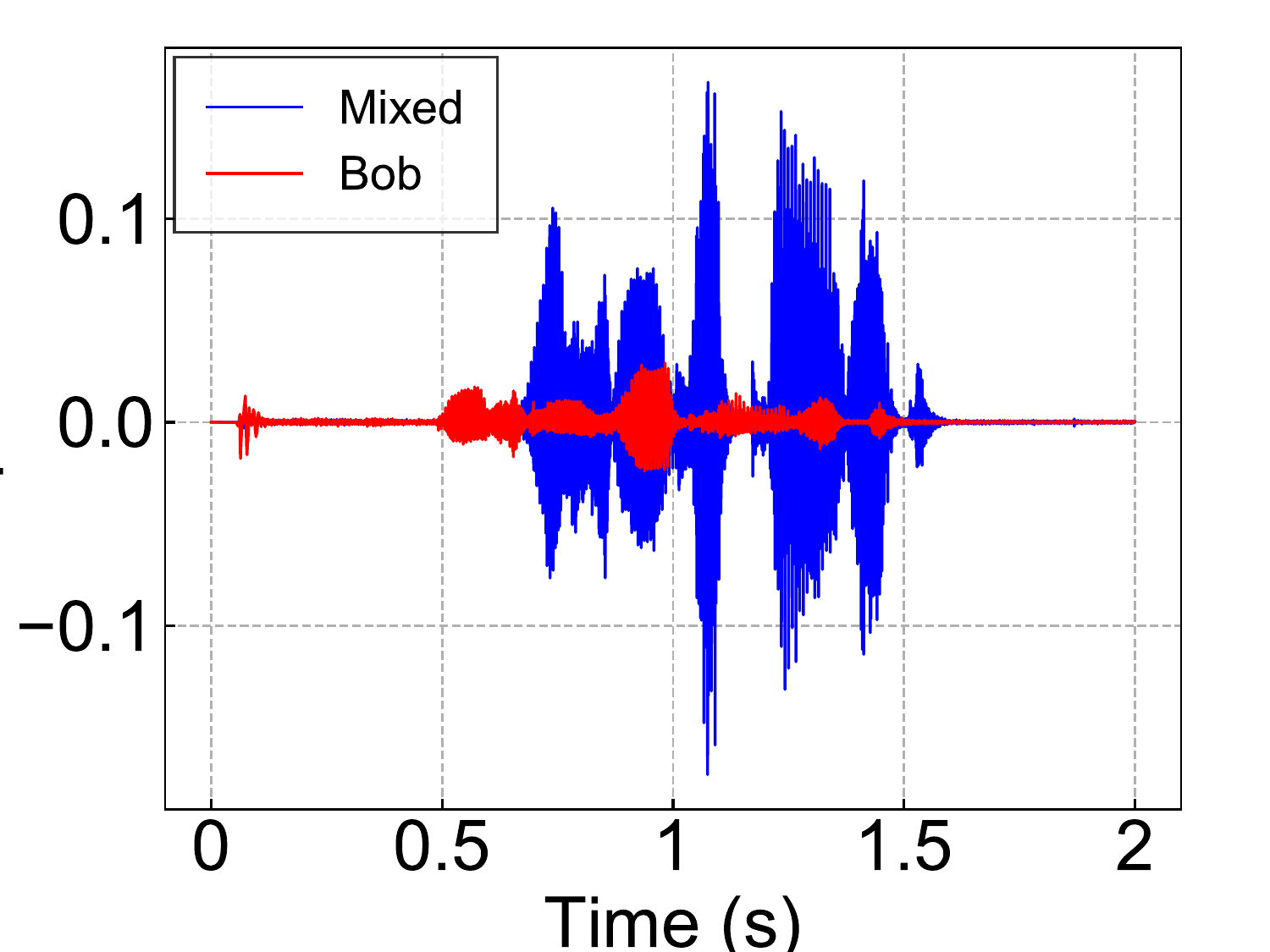}}
\subfigure[Distance = 2m]{\label{fig:c}\includegraphics[width=42mm]{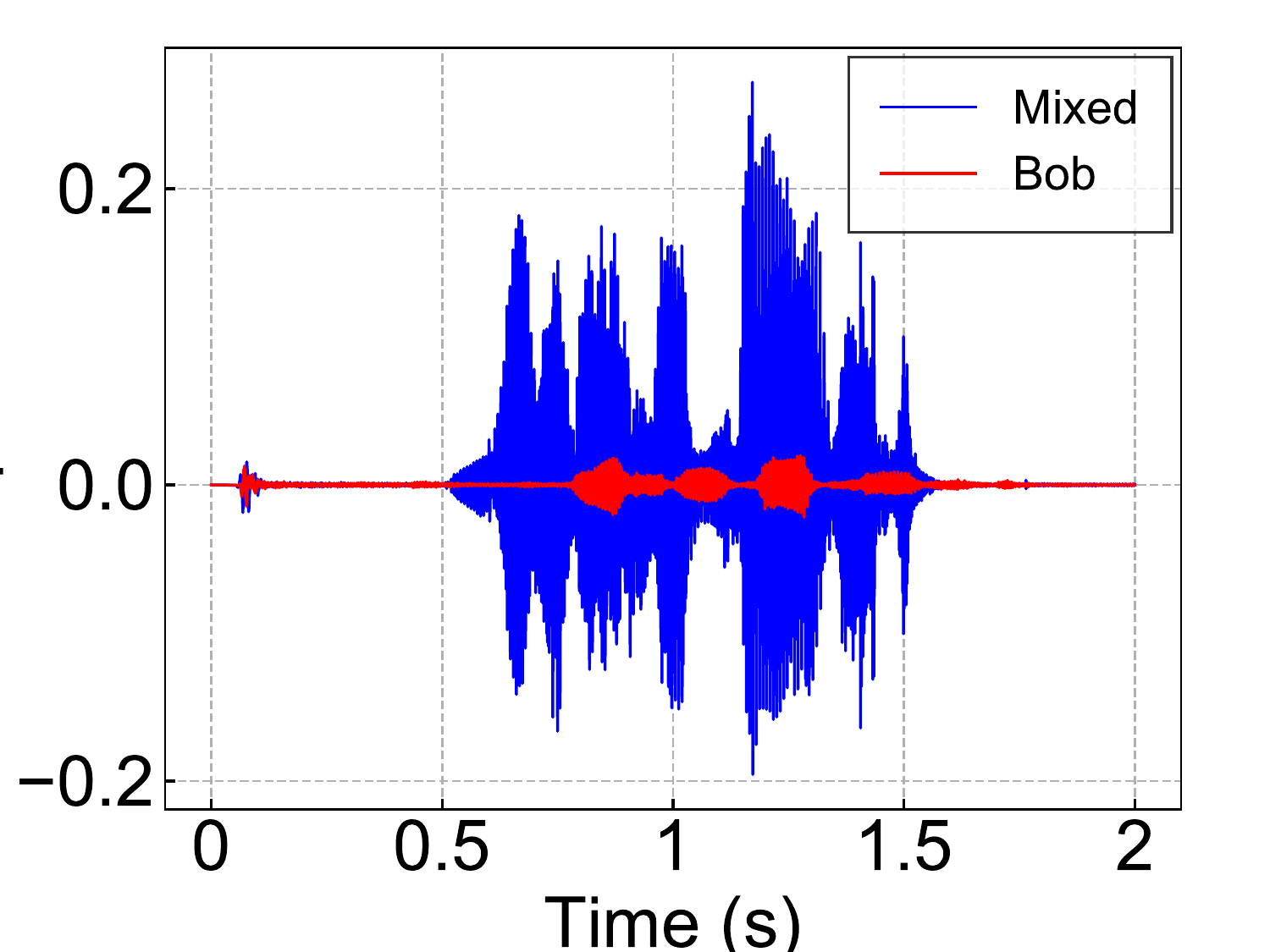}}
\subfigure[Distance = 3m]{\label{fig:d}\includegraphics[width=42mm]{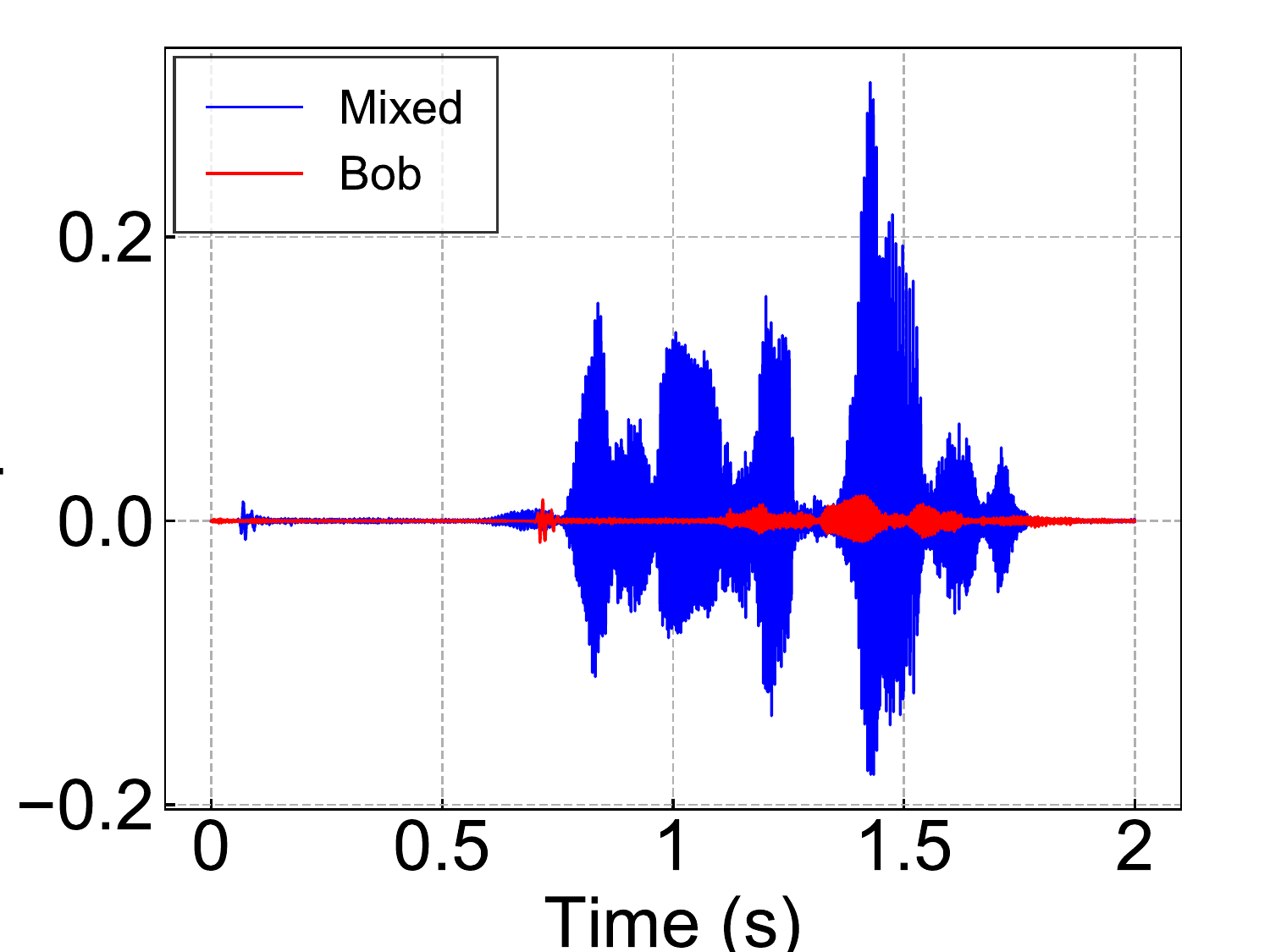}}
\vspace{-5pt}
\caption{Waveform of mixed audio and Bob's sole speech audio}
\vspace{-15pt}
\label{fig:disVSamp1}
\end{figure*}
\noindent
\textbf{User Case Study-2:} 
As depicted in Figure~\ref{fig:realworld}, in this user study, we evaluate how much of Bob's voice will be leaked to Alice's recorder with/without deploying \ours. We ask Bob and Alice to speak simultaneously, and also record Bob's sole speech with the same speech content. The mixed audio and Bob's individual speech audio are recorded by Alice's phone (Moto Z4), with varying distances between Alice and Bob (from 0.5 to 3 meters). 

\begin{figure}[h]
\centering     
\subfigure[Loudness  vs. Distance]{\label{fig:loud}\includegraphics[width=40mm]{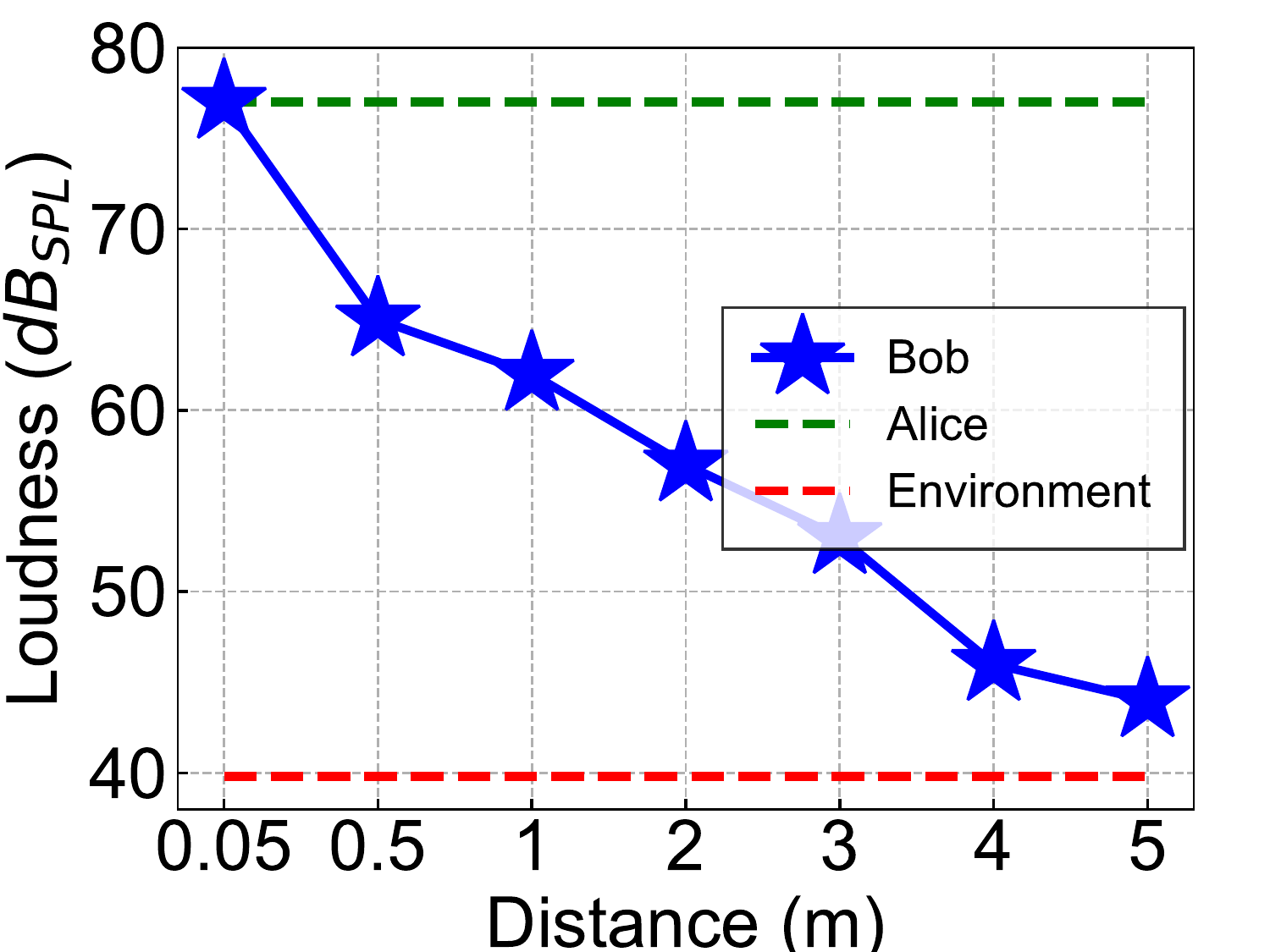}}
\subfigure[SONR vs. Distance]{\label{fig:snr}\includegraphics[width=40mm]{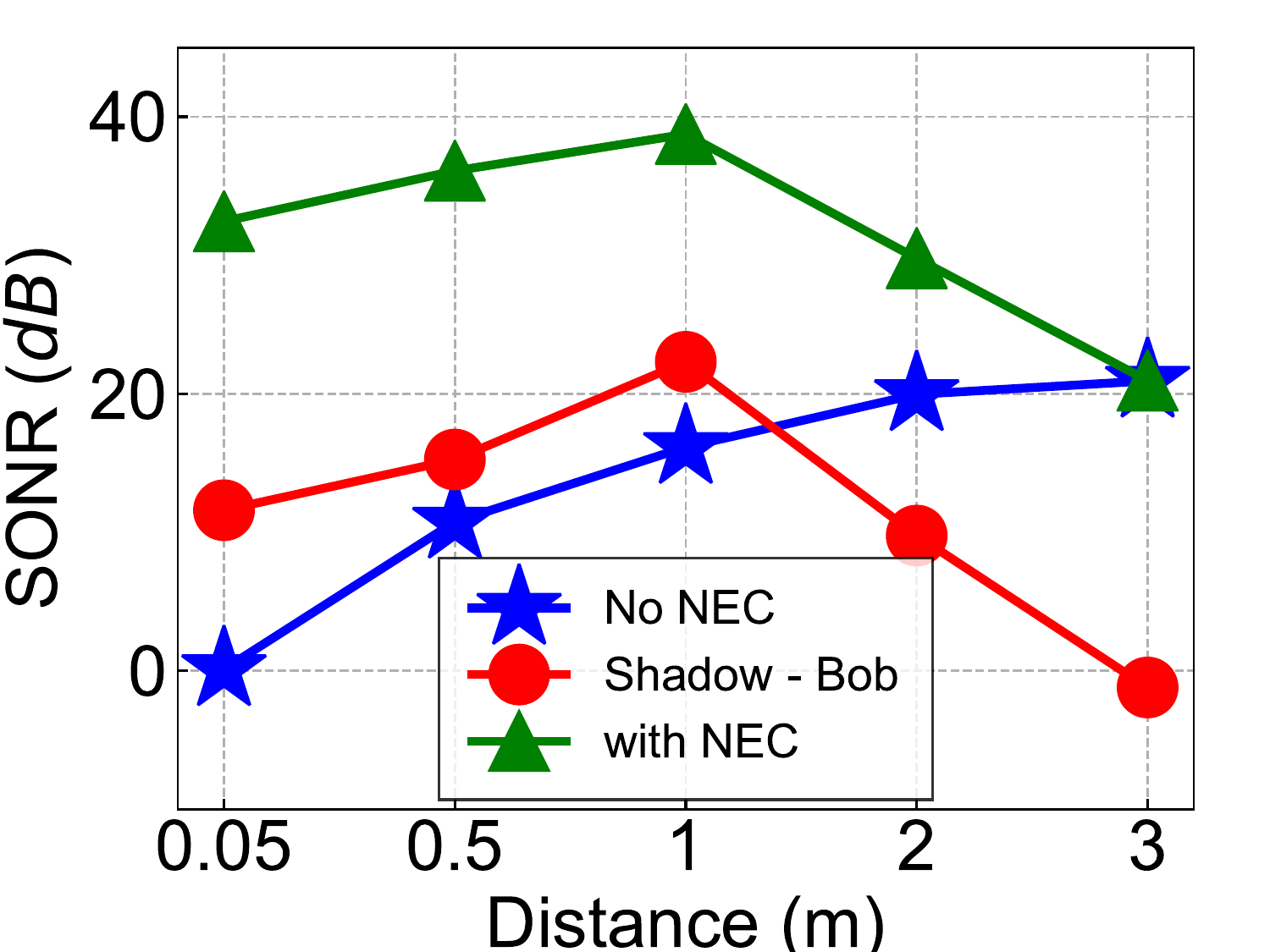}}
\vspace{-5pt}
\caption{Effectiveness of \ours across different distances}
\vspace{-10pt}
\end{figure}

Figure~\ref{fig:disVSamp1} visualizes the waveforms of Bob's audio and the mixed audio. We can see that with the increasing distance, Bob's audio contributes less to the mixed one. 
We further record Bob's sound pressure level (SPL) at Alice's position and present the result in Figure~\ref{fig:loud}. The result shows that the SPL of Bob's audio 
attenuates with the increasing distance, and its loudness  reaches $43 dB_{SPL}$ at the $5m$ distance (between Alice and Bob) with an environmental noise level of $39.8 dB_{SPL}$. 
In comparison, the SPL of Alice's voice recorded by her own recorder remains at $77 dB_{SPL}$.  
Given the large gap between the SPL of Alice and Bob's voices across different distances, and the attenuation of Bob's voice with the increasing distance, we can see that Bob only needs to cancel his voice over a short range (e.g., $2m$). 
Next, we further justify whether \ours can effectively overshadow Bob's sound across the distance.

Figure~\ref{fig:snr} presents the SONR results with/without NEC. 
When NEC is not deployed, 
the SONR between the recorded mixed audio and Bob's voice stays below 20dB, which implies that Bob's voice can be effectively captured by Alice's recorder. 
However, when Bob deploys \ours, even with a close distance ($<2m$), Bob's voice can be mostly overshadowed, with SONR reaching 30dB. 
As mentioned above, the strength of Bob's voice signals drops significantly beyond $2m$. 
Therefore, although the recorded shadow audio strength also degrades dramatically beyond $2m$, the effectiveness of \ours within $2m$ makes it a viable solution for target voice cancellation. 
\subsection{Comparison Study}\label{subsubsec-comparison-study}
Next, we perform a comparison experiment between \ours and \rev{two systems. The first one uses white noise to jam unauthorized recordings, which is commonly applied to commercial ultrasonic jammers. To simplify the jamming process, we manually add $10$dB white noise over the recording sound to simulate this type of jamming system. Notice that the volume of white noise is usually determined by different jammers, we use $10$dB based on our previous observation of the shadow sound volume on the same phone.
The second one is a scrambling-based voice hiding system called Patronus~\cite{li2020patronus}, which can hide the target recordings by scrambling with specially designed white noises and recover the target recordings at an authorized device. }
Given the mixed (joint) audios (e.g., two volunteers, one of which is target), we reproduce the scrambling algorithms of Patronus to hide a speaker's voice. 

\begin{figure}[h]
\centering
\subfigure[Hide Bob's voice.]{\includegraphics[width=0.24\textwidth]{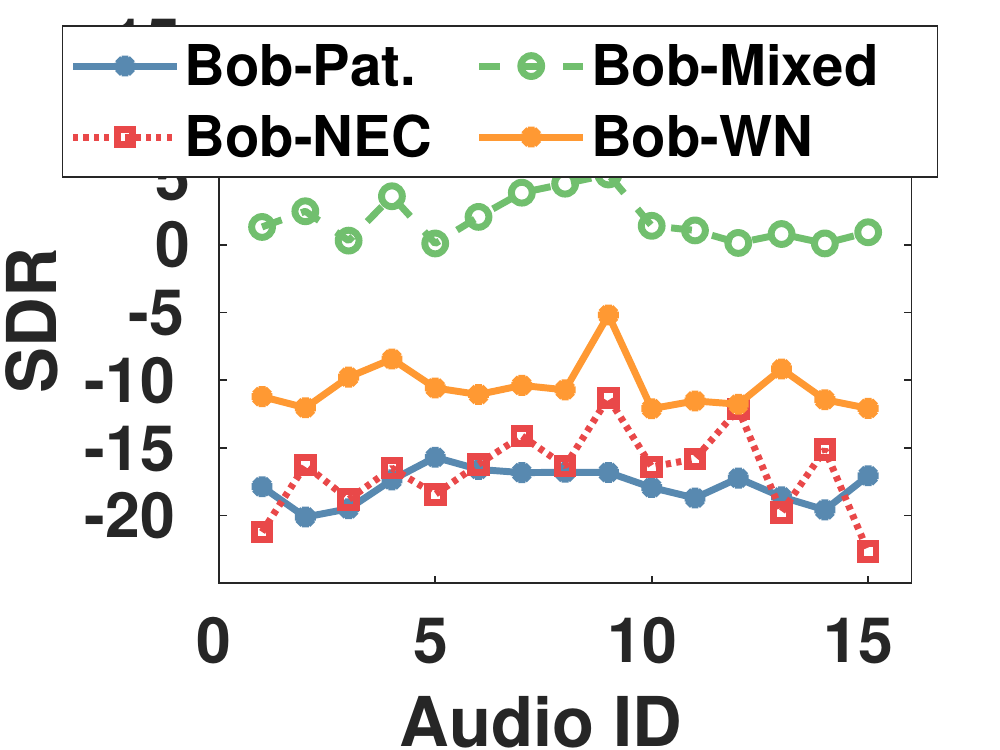}\label{fig:compare_hide}}
\subfigure[Retain Alice's voice.]{\includegraphics[width=0.24\textwidth]{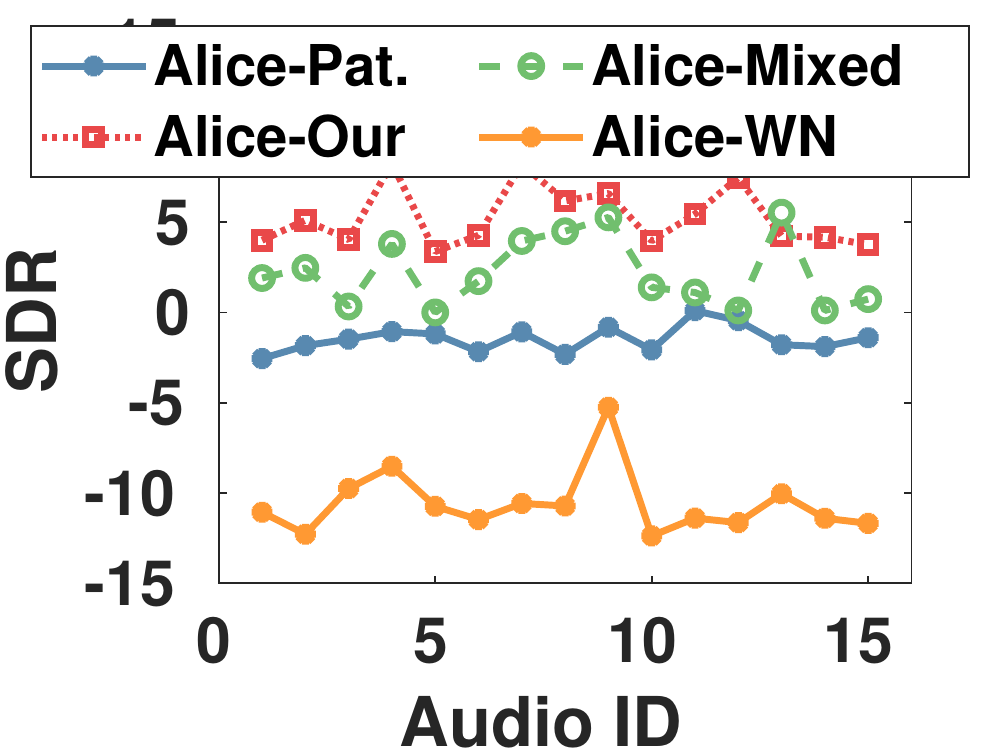}\label{fig:compare_focus}}
\vspace{-10pt}
\caption{Comparison study.}

\label{fig:comparison_and_person}
\end{figure}

We first compare the performance of voice hiding by computing the SDR of the target voice. 
\rev{Figure~\ref{fig:compare_hide} shows all of the three systems: \ours (Bob-NEC), White Noise (Bob-WN), and Patronus (Bob-Pat.) achieve a low SDR by effectively hiding the target voice in the mixed audio (Bob-Mixed). We find that, compared to \ours and Patronus, the white noise solution results in higher SDR, which means it retains more target voice than the other systems. 
Patronus and \ours can reduce the SDR of the mixed audios from $3$ dB to nearly $-20$ dB. Therefore, the voice hiding performance of \ours is on par with that of the specially designed scrambling-based Patronus, and better than the white noise scrambling approach.}
\rev{
Next, we evaluate the reception quality of Alice's voice in the presence of the three systems. 
As shown in Figure\ref{fig:compare_focus}, among the three systems, the White Noise approach cannot recover the disrupted voice, and therefore results in the lowest SDR for Alice's voice. For comparison, Patronus can recover a limited portion of scrambled sound by its recovery algorithm, and achieve low 
SDR for Alice's voice (i.e., $-2.5$ dB). The quality of Alice's voice after recovery is even lower than that of the raw mixed audios due to the influence of the scrambling noise. 
In comparison, \ours achieves a $5$ dB gain compared with the mixed audios in recovering Alice's voice, since \ours carefully nullifies Bob's voice in the mixed audio. }
This experiment result demonstrates that \ours could selectively hide a target speaker's voice without interfering with other speakers. Surprisingly, \ours can even improve the reception quality of others' recording.  

\vspace{-5pt}
\subsection{Running Time Analysis}\label{subsubsec-running-time-analysis}
To demonstrate the efficiency of our system, we measure the time consumption of each system module in Table~\ref{tbl:time}. Given 100 1s mixed audios, we evaluate the latency in two different hardware platforms: 1) desktop with a single NVIDIA 1080Ti GPU; 2) Raspberry Pi 4.
The total processing time of the DNN module in \ours is around $1.51$ ms, and the ultrasound modulation consumes $11.96$ ms on average, well below the lasting period of the $1$s chunks. In comparison, it takes 2.4$\times$ more time for VoiceFilter to process the same mixed audio. 
On the Raspberry Pi 4, the overhead of the selector is $293.7$ ms, which is faster than $446.2$ ms of VoiceFilter. 
The achieved latency $(<300)$ ms on the edge deployment using Pi 4 is less than the time offset tolerance of overshadowing, as discussed in $\S$\ref{subsubsec-latency-tolerance-for-overshadowing}, which further corroborates the feasibility of \ours. 

\vspace{-10pt}
\begin{table}[h]
\begin{center}
\caption{Time consumption of \ours with an audio sample lasting 1s}
\vspace{-8pt}
\begin{tabular}{c|c|ccc}
\hline
Platform &{System}&{Encoder}&{Selector}& {Broadcast}\\
\hline
\multirow{2}{*}{PC (1080Ti)} &\ours & 0.467ms &1.51ms & 11.96ms\\ 
\cline{2-5}
 & VoiceFilter~\cite{voiceFilter} & 0.467ms &3.65ms & 11.96ms\\ \hline 
\multirow{2}{*}{Rasp} & \ours & 12.7ms &293.7ms &11.96ms \\ \cline{2-5}
 & VoiceFilter~\cite{voiceFilter} & 12.7ms & 446.2ms& 11.96ms\\ \hline 
\end{tabular}
\vspace{-10pt}
\label{tbl:time}
\end{center}
\end{table}

\subsection{Parameter Study}\label{subsubsec-parameter-study}
\noindent
\textbf{Diversity of Hardware Dependence:}
The variance of the non-linearity for the hardware (e.g., microphones, amplifiers, filters) on smartphones can influence the optimal selection of the modulation parameters~\cite{SurfingAttack}, which in turn impacts the performance of our system.
Here, we evaluate our system using 7 different mobile devices listed in Table~\ref{tbl:phones}. Specifically, the carrier frequency $f_c$ is the dominant factor that affects the effectiveness of the non-linearity effect. All the tested smartphones have a range of acceptable frequency settings, and the best carrier frequency is listed in the brackets.  

\begin{table}[h]
\vspace{-5pt}
    \caption{Smartphones used for two user studies.}
    \vspace{-8pt}
    \label{tbl:phones}
    \centering
        \begin{tabular}{c|ccc}
        \hline
         \textbf{Model} & \textbf{Brand} &
        \textbf{Carrier $f_c$ (kHz)} & \textbf{Max Dis. (m)} \\
        \hline
        Moto Z4 & Motorola & 24-28 (28.0) & 3.2\\
        iPhone 7 P & Apple & 21-29 (27.8) & 0.49\\
        iPhone SE2 & Apple & 23-28 (25.2) & 1.77 \\
        iPhone X & Apple & 27-32 (25.3) & 0.43 \\
        iPad Air 3& Apple & 22-31 (28.0) & 3.72 \\
        Mi 8 Lite & Xiaomi & 24-32 (27.4) & 1.65 \\
        Pocophone & Xiaomi & 22-29 (26.3) & 0.7 \\
        Galaxy S9 & Samsung & 25-31 (27.2)  & 3.64\\
        \hline
        \end{tabular}
\end{table}

\noindent
\textbf{Diversity of Effective Distance:}
Our system can be deployed with various maximum effective distances with different smartphone recorders, ranging from 49 cm to 3.72 m, as shown in Table~\ref{tbl:phones}. The result also shows a great variance across recorders. We attribute this diversity to the difference in frequency response of these recorders, and the non-linearity of audio processing circuits.

\vspace{1mm}
\noindent
\textbf{Multiple Recorders:}
Since the performance of \ours can be affected by the variance of hardware, we investigate whether \ours system can be used to support multiple recorders simultaneously. To conduct this experiment, 
we use Moto Z4, Mi 8 Lite, POCOPHONE, and Galaxy S9 as recorders to eavesdrop on Bob's voice. With the collected recorded audios, we compute the SDR for recorded audios. For comparison, the SDR of mixed audio is also calculated to reveal the effect of \ours. We define that, if the SDR of recorded audio is less than the mixed audio, \ours is successfully performed. Our experiment result is presented in Table~\ref{tbl:more_recorders}. For three different carrier center frequency settings, we played 20 crafted mixed audios and run \ours to superpose shadow audio to affect three recorders' recording. The column named \textbf{1+}, \textbf{2+}, \textbf{3} means at least \textbf{1}, \textbf{2}, or \textbf{3} devices are affected simultaneously by \ours. And the reported values such as 20/20 denote that all the 20 recorded audios are unable to recognize Bob's voice. This result provides the supportive evidence that \ours is capable of operating in public and affecting multiple recorders by carefully tuning the system parameters.

\vspace{-10pt}
\begin{table}[h]
\caption{\ours's performance with multiple recorders. }
\vspace{-10pt}
\begin{center}
\begin{tabular}{c|c|ccc}
\hline
\multicolumn{2}{c|}{\textbf{Number of Recorder}} &{\textbf{1+}} & {\textbf{2+}}& {\textbf{3}}  \\
\hline
\multirow{3}{*}{\textbf{$f_c$ (kHz)}} & 26.3 &20/20 & 9/20 & 4/20 \\\cline{2-5}
& 27.2 & 20/20 & 15/20 & 11/20 \\\cline{2-5}
& 27.4 & 20/20 & 14/20 & 8/20 \\
\hline
\end{tabular}
\label{tbl:more_recorders}
\end{center}
\end{table}

%% file: 5_discussion.tex
\section{Discussion \& Limitation}\label{sec-discussion-and-limitation}
\noindent
\rev{\textbf{Limitation of non-linear effect:}
The success of \ours relies on the imperfection of the receivers' (e.g., Alice's) microphone. However, when the non-linear effect is not present due to two reasons: 1) the great precision of Alice's microphone or 2) the improper modulation parameter settings, our selective voice protection will no longer be effective.}

\noindent
\rev{\textbf{Limitation of protecting conversation:}
Although prior benchmark and user case experiments demonstrate that \ours can protect the target speaker's voice in the wild, it is a challenge to protect a conversation that involves multiple speakers while not disrupting other users (e.g., Alice). We failed to train a Selector model that is applicable to multiple target speakers with the current system architecture.
In future work, we will figure out how to integrate the multiple speakers' embeddings and re-design the Selector model to avoid removing Alice's voice in the private conversation.
}

\noindent
\rev{\textbf{Directional of Ultrasonic Speaker:}
In our prototype shown in Figure~\ref{fig:setup}, we assume the ultrasound speaker has the shadow audio ready before playing it. However, when we integrate the monitor, DNN models, and ultrasound speaker into one device and run it in a real-time manner, the shadow audio is dependent on the incoming mixed audio. In this case, one critical concern of \ours is whether the current mixed audio will be affected by the current shadow audio, therefore impacting the quality of future shadow audio. Fortunately, we can avoid it by putting the monitor and ultrasound speaker in \emph{opposite direction}. We find that by exploiting the directional property of the ultrasound speaker, the shadow audio is barely sensed by the \ours's monitor as it produces limited amplitude in its back direction. }

%% file: 6_conclusion.tex
\section{Conclusion}\label{sec-conclusion}
We present \ours, a lightweight AI-augmented voice protection system to protect the target speech without interfering with others' audio conversations.
As an end-to-end processing system, \ours first actively emits specially designed ultrasound signals to a recorder. 
Due to the non-linearity effect, a shadow sound is generated and superposed onto the received mixed sound at the recorder, which effectively cancels the target speaker's voice in the recordings.
To determine the frequency composition of the shadow sound, \ours leverages a tailored Deep Neural Network (DNN) to extract high-level speaker-specific but utterance-independent vocal features from the mixed sound.
By imitating the overshadowing in the air, we superpose the shadow audio  with the mixed audio in the training stage of the DNN model and deliver a one-fits-all model, which can be trained only once and deployed directly for new users.
Our experimental evaluations demonstrate \ours's efficacy in a wide variety of real-world scenarios. The results show that \ours effectively disables the microphones from recording the target speaker's voice.

%% file: 7_aknowledgment.tex
\section*{Acknowledgment}
\label{sec:ack}

We would like to thank our shepherd Michael Paulitsch and anonymous reviewers
for providing valuable feedback on our work. This work was supported in part by National Science Foundation grants CNS-1950171, CNS-1909177
and CCF-2007159.

%% file: refs.tex
\bibliographystyle{IEEEtran}
\bibliography{references}

%% file: main.bbl
\begin{thebibliography}{10}
\providecommand{\url}[1]{#1}
\csname url@samestyle\endcsname
\providecommand{\newblock}{\relax}
\providecommand{\bibinfo}[2]{#2}
\providecommand{\BIBentrySTDinterwordspacing}{\spaceskip=0pt\relax}
\providecommand{\BIBentryALTinterwordstretchfactor}{4}
\providecommand{\BIBentryALTinterwordspacing}{\spaceskip=\fontdimen2\font plus
\BIBentryALTinterwordstretchfactor\fontdimen3\font minus
  \fontdimen4\font\relax}
\providecommand{\BIBforeignlanguage}[2]{{%
\expandafter\ifx\csname l@#1\endcsname\relax
\typeout{** WARNING: IEEEtran.bst: No hyphenation pattern has been}%
\typeout{** loaded for the language `#1'. Using the pattern for}%
\typeout{** the default language instead.}%
\else
\language=\csname l@#1\endcsname
\fi
#2}}
\providecommand{\BIBdecl}{\relax}
\BIBdecl

\bibitem{li2020patronus}
L.~Li, M.~Liu, Y.~Yao, F.~Dang, Z.~Cao, and Y.~Liu, ``Patronus: preventing
  unauthorized speech recordings with support for selective unscrambling,'' in
  \emph{Proceedings of the 18th Conference on Embedded Networked Sensor Systems
  (SenSys)}, 2020, pp. 245--257.

\bibitem{scramble_noise}
Y.-C. Tung and K.~G. Shin, ``Exploiting sound masking for audio privacy in
  smartphones,'' in \emph{Proceedings of ACM Asia Conference on Computer and
  Communications Security}, 2019.

\bibitem{Jamming}
Y.~Chen, H.~Li, S.-Y. Teng, S.~Nagels, Z.~Li, P.~Lopes, B.~Y. Zhao, and H.~Z.
  0001, ``{Wearable Microphone Jamming.}'' \emph{CHI}, 2020.

\bibitem{li2020wihf}
C.~Li, M.~Liu, and Z.~Cao, ``Wihf: Enable user identified gesture recognition
  with wifi,'' in \emph{IEEE INFOCOM 2020-IEEE Conference on Computer
  Communications}.\hskip 1em plus 0.5em minus 0.4em\relax IEEE, 2020, pp.
  586--595.

\bibitem{li2021deep}
C.~Li, Z.~Cao, and Y.~Liu, ``Deep ai enabled ubiquitous wireless sensing: A
  survey,'' \emph{ACM Computing Surveys (CSUR)}, vol.~54, no.~2, pp. 1--35,
  2021.

\bibitem{guo2020deep}
H.~Guo, N.~Zhang, S.~Wu, and Q.~Yang, ``Deep learning driven wireless real-time
  human activity recognition,'' in \emph{ICC 2020-2020 IEEE International
  Conference on Communications (ICC)}.\hskip 1em plus 0.5em minus 0.4em\relax
  IEEE, 2020, pp. 1--6.

\bibitem{zhu2018indoor}
S.~Zhu, J.~Xu, H.~Guo, Q.~Liu, S.~Wu, and H.~Wang, ``Indoor human activity
  recognition based on ambient radar with signal processing and machine
  learning,'' in \emph{2018 IEEE international conference on communications
  (ICC)}.\hskip 1em plus 0.5em minus 0.4em\relax IEEE, 2018, pp. 1--6.

\bibitem{wang2018speaker}
Q.~Wang, C.~Downey, L.~Wan, P.~A. Mansfield, and I.~L. Moreno, ``Speaker
  diarization with lstm,'' in \emph{2018 IEEE International Conference on
  Acoustics, Speech and Signal Processing (ICASSP)}.\hskip 1em plus 0.5em minus
  0.4em\relax IEEE, 2018, pp. 5239--5243.

\bibitem{voiceFilter}
Q.~Wang, H.~Muckenhirn, K.~Wilson, P.~Sridhar, Z.~Wu, J.~R. Hershey, R.~A.
  Saurous, R.~J. Weiss, Y.~Jia, and I.~L. Moreno, ``Voicefilter: Targeted voice
  separation by speaker-conditioned spectrogram masking,'' in \emph{Proceedings
  of Interspeech}, 2019.

\bibitem{Spot_the_conversation_speaker_diarisation_in_the_wild}
J.~S. Chung, J.~Huh, A.~Nagrani, T.~Afouras, and A.~Zisserman, ``Spot the
  conversation: speaker diarisation in the wild,'' \emph{ArXiv}, 2020.

\bibitem{speaker_factor}
F.~{Castaldo}, D.~{Colibro}, E.~{Dalmasso}, P.~{Laface}, and C.~{Vair},
  ``Stream-based speaker segmentation using speaker factors and eigenvoices,''
  in \emph{Proceedings of IEEE International Conference on Acoustics, Speech
  and Signal Processing}, 2008.

\bibitem{i_vector_1}
W.~{Zhu} and J.~{Pelecanos}, ``Online speaker diarization using adapted
  i-vector transforms,'' in \emph{Proceedings of IEEE International Conference
  on Acoustics, Speech and Signal Processing (ICASSP)}, 2016.

\bibitem{i_vector_2}
M.~{Senoussaoui}, P.~{Kenny}, T.~{Stafylakis}, and P.~{Dumouchel}, ``A study of
  the cosine distance-based mean shift for telephone speech diarization,''
  \emph{IEEE/ACM Transactions on Audio, Speech, and Language Processing}, 2014.

\bibitem{fully_supervised_speaker_diarization}
A.~Zhang, C.~Wang, J.~Paisley, Q.~Wang, and Z.~Zhu, ``Fully supervised speaker
  diarization,'' in \emph{Proceedings of IEEE International Conference on
  Acoustics, Speech and Signal Processing (ICASSP)}, 2019.

\bibitem{visual_1}
A.~Ephrat, I.~Mosseri, O.~Lang, T.~Dekel, K.~Wilson, A.~Hassidim, W.~T.
  Freeman, and M.~Rubinstein, ``Looking to listen at the cocktail party: A
  speaker-independent audio-visual model for speech separation,'' \emph{ACM
  Trans. Graph.}, 2018.

\bibitem{visual_2}
T.~Afouras, J.~S. Chung, and A.~Zisserman, ``The conversation: Deep
  audio-visual speech enhancement,'' in \emph{INTERSPEECH}, 2018.

\bibitem{formants}
I.~R. Titze, R.~J. Baken, K.~W. Bozeman, S.~Granqvist, N.~Henrich, C.~T.
  Herbst, D.~M. Howard, E.~J. Hunter, D.~Kaelin, R.~D. Kent \emph{et~al.},
  ``Toward a consensus on symbolic notation of harmonics, resonances, and
  formants in vocalization,'' \emph{The Journal of the Acoustical Society of
  America}, 2015.

\bibitem{caField}
C.~Yan, Y.~Long, X.~Ji, and W.~Xu, ``The catcher in the field: A fieldprint
  based spoofing detection for text-independent speaker verification,'' in
  \emph{Proceedings of ACM CCS}, 2019.

\bibitem{timbre}
F.~Winckel and T.~Binkley, ``Music, sound and sensation : a modern
  exposition,'' 1967.

\bibitem{LTAS}
A.~L{\"o}fqvist and B.~Mandersson, ``{Long-time average spectrum of speech and
  voice analysis.}'' \emph{Folia phoniatrica}, 1987.

\bibitem{speech_principal}
A.~Jongman, ``Acoustics of american english speech: A dynamic approach,''
  \emph{Language and Speech}, 1995.

\bibitem{MUTE}
S.~Shen, N.~Roy, J.~Guan, H.~Hassanieh, and R.~R. Choudhury, ``Mute: Bringing
  iot to noise cancellation,'' in \emph{Proceedings of ACM SIGCOMM}, 2018.

\bibitem{10}
L.~{Wan}, Q.~{Wang}, A.~{Papir}, and I.~L. {Moreno}, ``Generalized end-to-end
  loss for speaker verification,'' in \emph{Proceedings of IEEE International
  Conference on Acoustics, Speech and Signal Processing (ICASSP)}, 2018.

\bibitem{fleischer2015formant}
M.~Fleischer, S.~Pinkert, W.~Mattheus, A.~Mainka, and D.~M{\"u}rbe, ``Formant
  frequencies and bandwidths of the vocal tract transfer function are affected
  by the mechanical impedance of the vocal tract wall,'' \emph{Biomechanics and
  modeling in mechanobiology}, vol.~14, no.~4, pp. 719--733, 2015.

\bibitem{igras2013length}
M.~Igras, B.~Zi{\'o}{\l}ko, and M.~Zi{\'o}{\l}ko, ``Length of phonemes in a
  context of their positions in polish sentences,'' in \emph{2013 International
  Conference on Signal Processing and Multimedia Applications (SIGMAP)}.\hskip
  1em plus 0.5em minus 0.4em\relax IEEE, 2013, pp. 59--64.

\bibitem{trauzettel2012standardized}
S.~Trauzettel-Klosinski and K.~Dietz, ``Standardized assessment of reading
  performance: The new international reading speed texts irest,''
  \emph{Investigative ophthalmology \& visual science}, 2012.

\bibitem{BackDoor}
N.~Roy, H.~Hassanieh, and R.~Roy~Choudhury, ``Backdoor: Making microphones hear
  inaudible sounds,'' in \emph{Proceedings of ACM MobiSys}, 2017.

\bibitem{SurfingAttack}
Q.~Yan, K.~Liu, Q.~Zhou, H.~Guo, and N.~Zhang, ``Surfingattack: Interactive
  hidden attack on voice assistants using ultrasonic guided wave,'' in
  \emph{Proceedings of Network and Distributed Systems Security (NDSS)
  Symposium}, 2020.

\bibitem{DolphinAttack}
G.~Zhang, C.~Yan, X.~Ji, T.~Zhang, T.~Zhang, and W.~Xu, ``Dolphinattack:
  Inaudible voice commands,'' in \emph{Proceedings of ACM CCS}, 2017.

\bibitem{yang2019hiding}
H.~Yang, S.~Bae, M.~Son, H.~Kim, S.~M. Kim, and Y.~Kim, ``Hiding in plain
  signal: Physical signal overshadowing attack on {LTE},'' in \emph{Proceedings
  of {USENIX} Security Symposium ({USENIX} Security)}, 2019.

\bibitem{SDR}
E.~{Vincent}, R.~{Gribonval}, and C.~{Fevotte}, ``Performance measurement in
  blind audio source separation,'' \emph{IEEE Transactions on Audio, Speech,
  and Language Processing}, 2006.

\bibitem{Amplifier}
A.~Bioacoustics, ``Portable ultrasonic power amplifier,'' in
  \emph{http://www.avisoft.com/playback/power-amplifier/}, Retrieved by July
  19th 2021.

\bibitem{Vifa}
``Ultrasonic dynamic speaker vifa,'' in
  \emph{http://www.avisoft.com/playback/vifa/}, Accessed on April 13, 2022.

\bibitem{panayotov2015librispeech}
V.~Panayotov, G.~Chen, D.~Povey, and S.~Khudanpur, ``Librispeech: an asr corpus
  based on public domain audio books,'' in \emph{Proceedings of IEEE
  International Conference on Acoustics, Speech and Signal Processing
  (ICASSP)}, 2015.

\bibitem{varga1993assessment}
A.~Varga and H.~J. Steeneken, ``Assessment for automatic speech recognition:
  Ii. noisex-92: A database and an experiment to study the effect of additive
  noise on speech recognition systems,'' \emph{Speech communication}, 1993.

\bibitem{Widar3.0}
Y.~Zheng, Y.~Zhang, K.~Qian, G.~Zhang, Y.~Liu, C.~Wu, and Z.~Yang,
  ``Zero-effort cross-domain gesture recognition with wifi,'' in
  \emph{Proceedings of ACM MobiSys}, 2019.

\bibitem{CrossSense}
J.~Zhang, Z.~Tang, M.~Li, D.~Fang, P.~Nurmi, and Z.~Wang, ``Crosssense: Towards
  cross-site and large-scale wifi sensing,'' in \emph{Proceedings of ACM
  MobiCom}, 2018.

\bibitem{WER}
{Ye-Yi Wang}, A.~{Acero}, and C.~{Chelba}, ``Is word error rate a good
  indicator for spoken language understanding accuracy,'' in \emph{Proceedings
  of IEEE Workshop on Automatic Speech Recognition and Understanding}, 2003.

\end{thebibliography}
